\documentclass[aps,prx,reprint, superscriptaddress]{revtex4-1}
\usepackage[english]{babel}
\usepackage{amsmath}
\usepackage{amsfonts}
\usepackage{amssymb}
\usepackage{graphicx}
\usepackage{bm}
\usepackage{bbm}
\usepackage{braket}
\usepackage{color, ulem}
\AtBeginDocument{%
	\newwrite\bibnotes
	\def\bibnotesext{Notes.bib}
	\immediate\openout\bibnotes=\jobname\bibnotesext
	\immediate\write\bibnotes{@CONTROL{REVTEX41Control}}
	\immediate\write\bibnotes{@CONTROL{%
			apsrev41Control,author="08",editor="1",pages="1",title="0",year="1"}}
	\if@filesw
	\immediate\write\@auxout{\string\citation{apsrev41Control}}%
	\fi
}%

\begin{document}
\title{Site-selective quantum control in an isotopically enriched $^{28}$Si/SiGe quadruple quantum dot}

\author{A. J. Sigillito}
\affiliation{Department of Physics, Princeton University, Princeton, New Jersey 08544, USA}
\author{J. C. Loy}
\affiliation{Department of Physics, Princeton University, Princeton, New Jersey 08544, USA}
\author{D. M. Zajac}
\affiliation{Department of Physics, Princeton University, Princeton, New Jersey 08544, USA}
\author{M. J. Gullans}
\affiliation{Department of Physics, Princeton University, Princeton, New Jersey 08544, USA}
\author{L. F. Edge}
\affiliation{HRL Laboratories LLC, 3011 Malibu Canyon Road, Malibu, California 90265, USA}
\author{J. R. Petta}
\affiliation{Department of Physics, Princeton University, Princeton, New Jersey 08544, USA}

\pacs{03.67.Lx, 73.63.Kv, 85.35.Gv}

\renewcommand{\topfraction}{0.9}	
\renewcommand{\bottomfraction}{0.8}	
\setcounter{topnumber}{2}
\setcounter{bottomnumber}{2}
\setcounter{totalnumber}{4}     
\setcounter{dbltopnumber}{2}    
\renewcommand{\dbltopfraction}{0.7}	
\renewcommand{\textfraction}{0.027}	
\renewcommand{\floatpagefraction}{0.8}	
\renewcommand{\dblfloatpagefraction}{0.7}	

\begin{abstract}
Silicon spin qubits are a promising quantum computing platform offering long coherence times, small device sizes, and compatibility with industry-backed device fabrication techniques. In recent years, high fidelity single-qubit and two-qubit operations have been demonstrated in Si. Here, we demonstrate coherent spin control in a quadruple quantum dot fabricated using isotopically enriched $^{28}$Si. We tune the ground state charge configuration of the quadruple dot down to the single electron regime and demonstrate tunable interdot tunnel couplings as large as 20 GHz, which enables exchange-based two-qubit gate operations. Site-selective single spin rotations are achieved using electric dipole spin resonance in a magnetic field gradient. We execute a resonant-CNOT gate between two adjacent spins in 270 ns.
\end{abstract}

\maketitle

Quantum processors based on spins in semiconductors \cite{loss1998,hanson2007spins, zwanenburg2013silicon} are rapidly becoming a strong contender in the global race to build a quantum cofmputer. In particular, silicon is an excellent host material for spin-based quantum computing by virtue of its small spin-orbit coupling and long spin coherence times \cite{tyryshkin2012electron, veldhorst2014addressable}. Within the past few years, tremendous progress has been made in achieving high fidelity single-qubit \cite{Yoneda2018, Yang2018} and two-qubit control \cite{Veldhorst2015,Zajac2018,Watson2018,Huang2018, Xue2018} in silicon. Scalable one-dimensional arrays of silicon quantum dots have been demonstrated \cite{ZajacScalable}, and in GaAs, where electron wavefunctions are comparably large, both one- \cite{Noiri2016,Otsuka2016,Ito2018} and two-dimensional arrays \cite{Mortemousque2018, Udi2018} of spins have been fabricated. Despite this progress, quantum control of spins in silicon has been limited to one- and two-qubit devices. Scaling beyond two-qubit devices opens the door to important experiments which are currently out of reach, including error correction \cite{Reedec,Schindlerec}, quantum simulation \cite{Hensgens2017,GeorgescuQS,Barthelemyqs,Byrnesqs}, and demonstrations of time crystal phases \cite{Barnestc}.

In this Letter, we demonstrate operation of a four-qubit device fabricated using an isotopically enriched $^{28}$Si/SiGe heterostructure. The device offers independent control of all four qubits, as well as pairwise two-qubit gates mediated by the exchange interaction \cite{petta2005}. We demonstrate control and measurement of the charge state of the array, and operate in the regime where each dot contains only one electron. We perform electric dipole spin resonance (EDSR) spectroscopy on all four qubits to show that they have unique spin resonance frequencies. Finally, we modulate the tunnel coupling between adjacent dots and demonstrate a resonant-CNOT gate \cite{Zajac2018,Russ2018}.

\begin{figure}
	\begin{center}
		\includegraphics[width=\columnwidth]{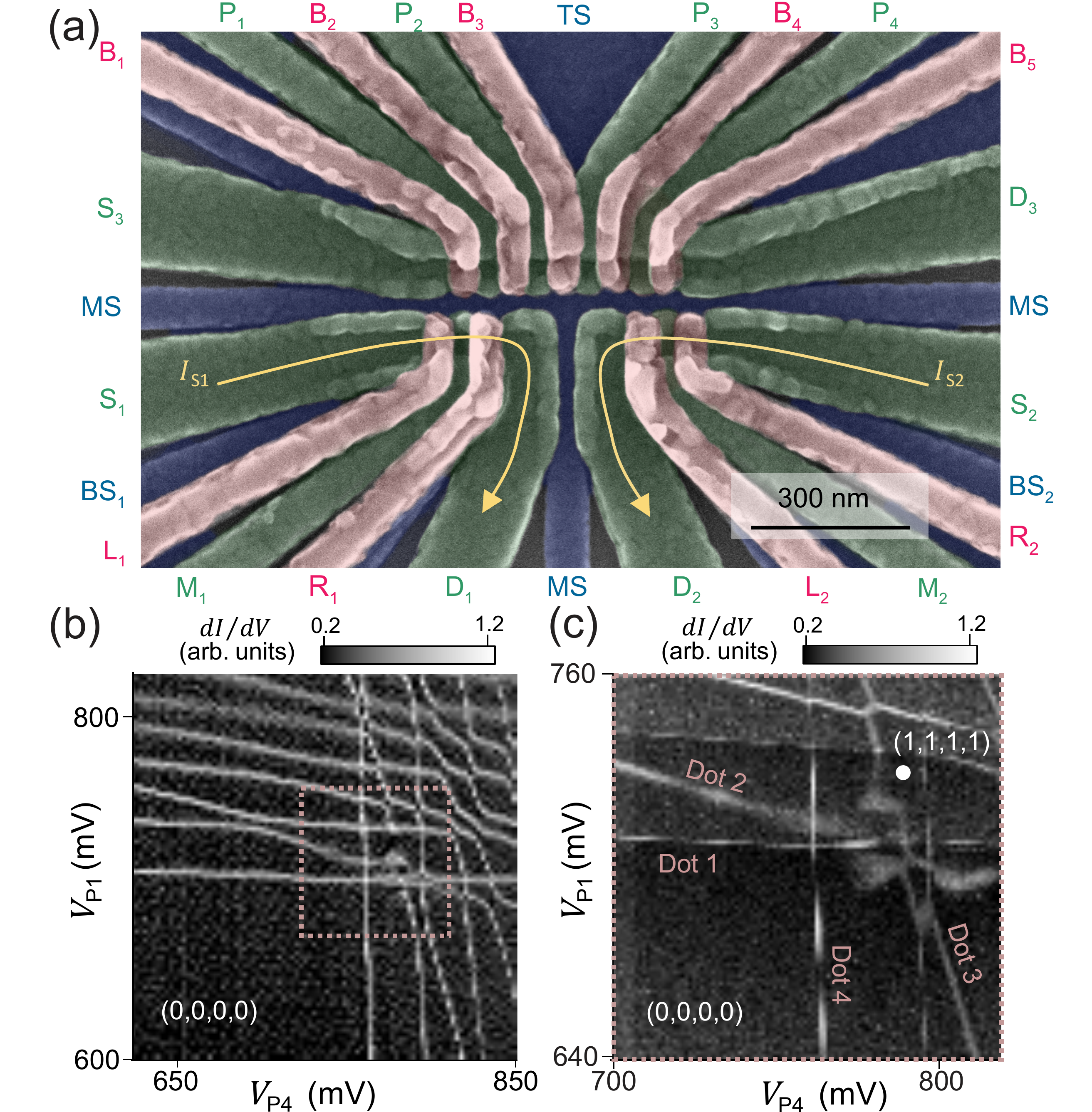}
		\caption{
			(a) False-color scanning electron micrograph of the device. The qubit electrons are accumulated underneath the plunger gates $P_{i}$ and the charge sensor dots are formed under the gates $M_{i}$. (b) A large-scale charge stability diagram shows that the array can be emptied of free electrons to reach the (0,0,0,0) charge state. The sensing signal $dI/dV$ is obtained by combining the differentiated signal from both charge detectors. (c) Charge stability diagram acquired near the $(1,1,1,1)$ charge state, where quantum control is performed.}
		\label{fig1}
	\end{center}	
	\vspace{-0.6cm}
\end{figure}

Four spin qubits are arranged in a linear array using an overlapping gate architecture, as shown in Fig.~1(a) \cite{zajac2015reconfigurable}. Single spin qubits are formed by accumulating a electron under each plunger gate: $P_1$, $P_2$, $P_3$, and $P_4$. The couplings between dots and between dots and the charge reservoirs formed beneath gates $S_3$ and $D_3$ are tuned by adjusting the barrier gate voltages $V_{Bi}$. Charge sensing is performed by monitoring the currents $I_{S1}$ and $I_{S2}$ through two proximal quantum dot charge detectors located in the lower half of the device. 

Charge stability diagrams for the array are shown in Figs.~1(b-c). To obtain good charge sensitivity for all four dot charge transitions $dI/dV$ = $dI_{S1}/dV_{P1} + dI_{S1}/dV_{P4} + dI_{S2}/dV_{P1} + dI_{S2}/dV_{P4}$  is plotted as a function of $V_{P1}$ and $V_{P4}$. In Fig.~1(b) we show that we can achieve the $(N_{1},N_{2}, N_{3}, N_{4}) = (0,0,0,0)$ charge state, where $N_{i}$ denotes the number of electrons in dot $i$. The $(0,0,0,0)$ charge state is evident from the large region devoid of charge transitions in the lower left corner of the figure. The device is typically operated in the (1,1,1,1) charge state, which is labeled in the zoomed-in charge stability diagram of Fig.~1(c). The capacitive coupling between plunger gates and neighboring dots (i.e.\ gate P$_{1}$ and dot 2) is naturally an order of magnitude weaker than the coupling between a plunger gate and the dot formed directly underneath it (i.e.\ gate P$_{1}$ and dot 1) \cite{zajac2015reconfigurable}. It can therefore be challenging to distinguish charge transitions in adjacent dots (P$_{1}$ and P$_{2}$, or P$_{3}$ and P$_{4}$) in two-dimensional charge stability plots since the slopes are very similar. To more clearly distinguish the charge transitions, an artificial cross-coupling was added between each plunger gate and its neighboring plunger gates in software (details available in the supplementary information \cite{SOM}), such that a sweep of $V_{P1}$ and   $V_{P4}$ induces transitions not only in dots 1 and 4, but also in dots 2 and 3. With the (1,1,1,1) charge state having been identified, we next establish virtual gates, which significantly streamline device tuning. 

Virtual gates have been described in detail and compensate for the effects of device cross-capacitance through software corrections that effectively invert the capacitance matrix \cite{Mills2018, baart2016single, Hensgens2017, vanDiepen2018, Volk2019}. Whenever the voltage of gate $V_{Pi}$ is adjusted to tune the chemical potential of dot $i$, the voltages on adjacent gates $V_{P(i-1)}$ and $V_{P(i+1)}$ are modified by a calibrated amount to keep the chemical potentials of dots $i-1$ and $i+1$ constant. The measured capacitance matrix is given in the supplementary information \cite{SOM} and is used to establish the virtual gate space \cite{Mills2018}.

Pairwise charge stability diagrams measured using virtual gates are shown in Figs.~2(a--c). As the two virtual gates $u_i$ and $u_{i+1}$ are swept, the charge sensor currents $I_{S1}$ and $I_{S2}$ are measured. Here we plot $I_{S1}$ - $I_{S2}$ as it results in higher charge sensing contrast. Gates not being swept [i.e.\ $u_3$ and $u_4$ in Fig.~2(a)] are held fixed at the same chemical potential as the source and drain reservoirs to enable fast loading and unloading of electrons throughout the array. The orthogonality of the charge transitions indicates that we have independent control of each quantum dot's chemical potential.

\begin{figure}
	\includegraphics[width=\columnwidth]{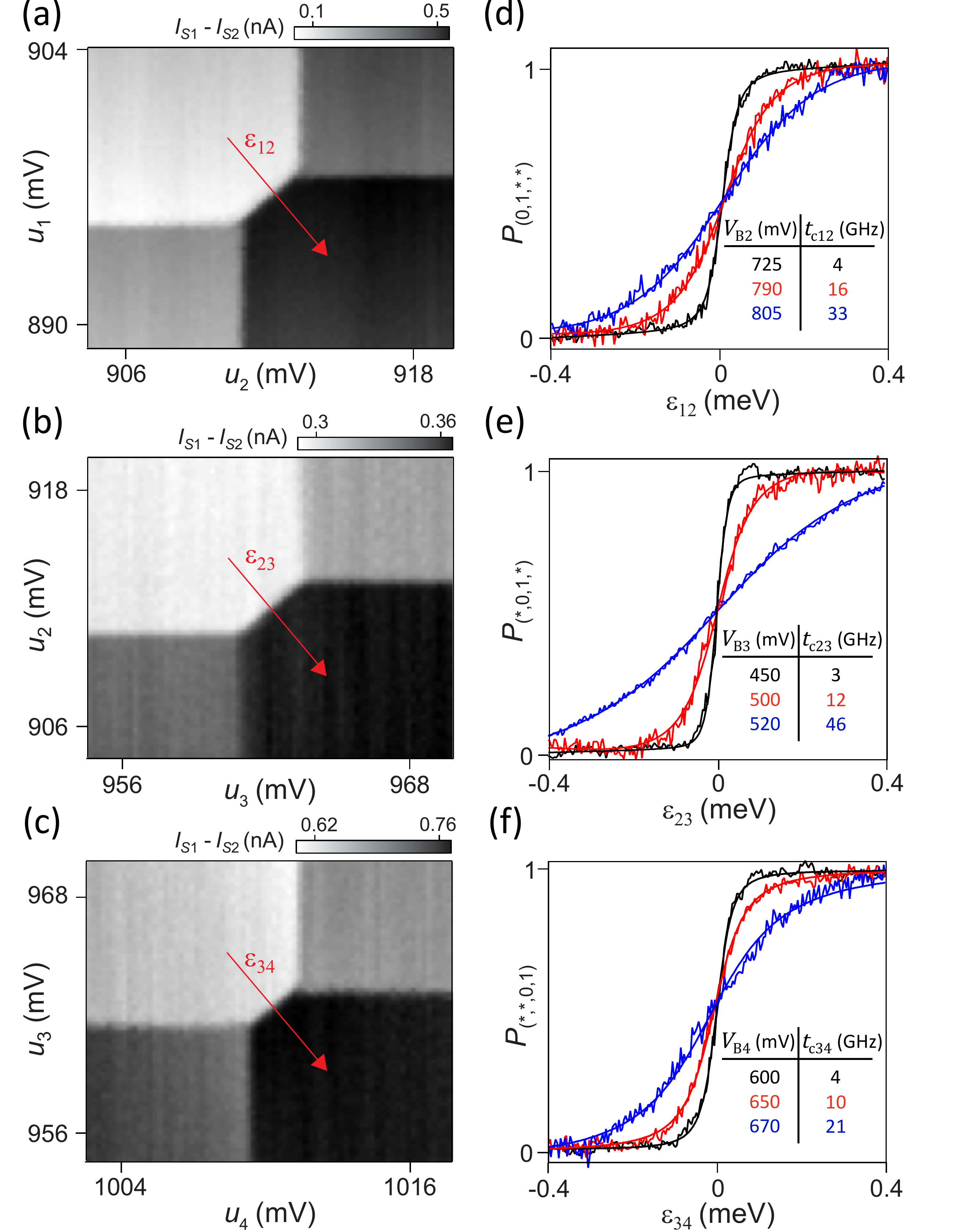}
	\caption{	
		Pairwise charge stability diagrams measured for (a) dots 1 and 2, (b) dots 2 and 3, and (c) dots 3 and 4. Here we use virtual gates to independently tune each dot's chemical potential. (d--f) Charge state occupation measured as a function of detuning $\epsilon_{ij}$ for various barrier gate voltages $V_{Bi}$. An increase in interdot tunnel coupling broadens the interdot charge transitions. The data are fit to theory (solid lines) to extract the interdot tunnel coupling $t_{cij}$ between dots $i$ and $j$ \cite{dicarlo2004}.
	}
	\label{fig2}
\end{figure}

Loss \& DiVincenzo suggested that the exchange interaction between two spins could be modulated by adjusting the height of the tunnel barrier separating the spins \cite{loss1998}. We demonstrate control over all of the interdot tunnel couplings ($t_{cij}$) in our device by measuring the charge state occupation as a function of detuning $\epsilon_{ij}$ for different barrier gate voltages $V_{Bi}$. Figure 2(d) plots $P_{(0,1,*,*)}$ along the detuning axis shown in Fig.\ 2(a). The * denotes that the chemical potential of the dot is held at the same value as the source and drain chemical potentials. As the tunnel coupling is increased, the charge delocalizes across adjacent dots and the interdot charge transition broadens. These data are fit as described in \cite{dicarlo2004, Petta2003} to extract the interdot tunnel coupling. The lever arm conversion between gate voltage and energy is determined by measuring finite bias triangles for each pair of dots as reported in Table 1. As demonstrated by the data, the device offers a high degree of control, with tunnel coupling tunable  from $2t_{cij} \approx k_BT_e \approx$ 2 GHz to many 10's of GHz. Here $k_B$ is Boltzmann's constant and the electron temperature $T_e$ $\approx$ 90 mK is determined by fitting the charge transitions to the source and drain to a
Fermi function as described in \cite{ZajacScalable}.

Site-selective single spin rotations are achieved using EDSR \cite{pioro2008electrically, Tokura2006} in the presence of a magnetic field gradient generated by a Co micromagnet \cite{SOM}. The field from the micromagnet $B_{i}^{M}$  is different at each dot and therefore each spin has a unique electron spin resonance frequency $f_i$ given by  $hf_{i} = g\mu_B(B_{ext}+B_{i}^{M})$, where $h$ is Planck's constant, $g \approx 2$ is the Land\'e g factor, and $B_{ext}$ is the externally applied magnetic field. Our micromagnet design is similar to that used by Yoneda \textit{et al.} \cite{Yoneda2015}, but it has a slanting edge (as seen from above) that extends the field gradient over the entire quadruple quantum dot \cite{SOM}.

To map out $B_{i}^{M}$, EDSR spectroscopy is performed on each qubit. During spectroscopy, the array is configured  such that only dot $i$ contains a single electron -- all other dots are empty. A frequency chirped microwave pulse ($\pm$ 15 MHz around a frequency $f$ for 120 $\mu$s) is applied to gate MS. If the chirped pulse sweeps through the spin resonance frequency $f_i$ of dot $i$, its spin will end up in a mixed state. Here our chirped pulses are not adiabatic, but are nonetheless convenient for identifying spin resonance conditions, as the linewidth of electron spins in $^{28}$Si can be narrow ($<$100 kHz). The spin state of dot $i$ is then measured through spin-selective tunneling to the leads \cite{elzerman2004single}. In the case of $i=2$ and $i=3$, where the dots are not directly connected to the leads, the electron is shuttled to the edge of the array and read out in dots 1 or 4, respectively \cite{Mills2018}. Loading the array follows the read out sequence in reverse. These measurements are repeated over a range of $B_{ext}$ spanning 250 -- 450 mT. The spectra for all four qubits are summed and plotted in Fig.~3(a). The field gradient from the micromagnet separates the qubit resonance frequencies by hundreds of MHz, as highlighted by the linecut through the data in Fig.~3(b). For comparison, in silicon devices relying on Stark shifts or interface disorder for spin selectivity, the qubit splitting is typically a few 10's of MHz \cite{Veldhorst2015,Huang2018}.

\begin{table} [t]
	\begin{center}
		\begin{tabular}{c | c | c | c| c | c }
			Dot & $\alpha$~(meV/mV) & E$_\text{c}$~(meV)& $B^{M}_{i}$ (mT) & $T_{2}^*$ ($\mu$s)& $T_{2,  echo}$ ($\mu$s) \\
			\hline\hline
			1 & 0.14 & 4.5 & 137.6 & 2.6& 41\\
			2 & 0.13 & 4.7 & 165.8 & 1.5& 31 \\
			3 & 0.14 & 4.5 & 194.3 & 10.4& 72 \\
			4 & 0.15 & 4.7 & 199.2 & 9.4& 109 \\
			
		\end{tabular}
		\caption{
			Summary of single-qubit parameters including lever-arm conversion between gate voltage and energy $\alpha$, charging energy $E_{\rm c}$, magnetic field offset due to the micromagnet $B^{M}_{i}$, spin dephasing time $T_{2}^*$, and spin coherence time $T_{2, echo}$.
		}
		\label{table1}
	\end{center}
\end{table}

We next measure the spin dephasing times $T_{2}^*$ and spin coherence times $T_2$ for each qubit (see Table 1).  In natural silicon, spin coherence is limited by the hyperfine interaction with the 4.7\% abundant $^{29}$Si nuclei. Here, our device consists of an isotopically enriched $^{28}$Si quantum well, which is 4.9 nm thick and has only an 800 ppm residual concentration of $^{29}$Si \cite{Deelman2016,Richardson2017}. The buffer layers, however, consist of $^{nat}$Si and $^{nat}$Ge containing residual spin-1/2 and spin-9/2 nuclei, respectively. Wavefunction overlap with these nuclei will be non-negligible given the relatively thin quantum well \cite{Witzel2012}.

$T_2^*$ is determined through measurements of Ramsey fringes. For each Ramsey decay curve the data are integrated over 15 minutes. Qubits 3 and 4 show a nearly tenfold increase in $T_{2}^*$ compared with  $T_{2}^*$ $\sim$ 1 $\mu$s for electron spins in natural silicon, whereas qubits 1 and 2 have a dephasing time that is comparable to natural silicon. A simple Hahn echo pulse sequence significantly extends the coherence times, as summarized in Table.~1. Similar fluctuations in the coherence times have been observed in other devices \cite{Watson2018,Zajac2018,Huang2018}  and may be due to sampling over a relatively small number of spin-carrying nuclei in the quantum well and SiGe barrier layers. Another reason for the fast dephasing in qubits 1 and 2 could be charge noise which has been shown to shorten $T_1$ \cite{Borjans2018} and $T_2$ \cite{Yoneda2018} in the presence of field gradients and is discussed in the supplementary information \cite{SOM}. Due to the wedge shaped geometry of the micromagnet, the field gradient experienced by qubits 1 and 2 is significantly larger than at sites 3 and 4 which is evident from the large change in field offsets $B_{i}^{M}$ between dots 1-3, and a relatively small change in $B_{i}^{M}$ between dots 3 and 4 (see Table~1). The short coherence times in dots 1 and 2 are still comparable with the times reported in natural Si/SiGe and are not prohibitive for two-qubit operation.

\begin{figure}[t]
	\begin{center}
		\includegraphics[width=\columnwidth]{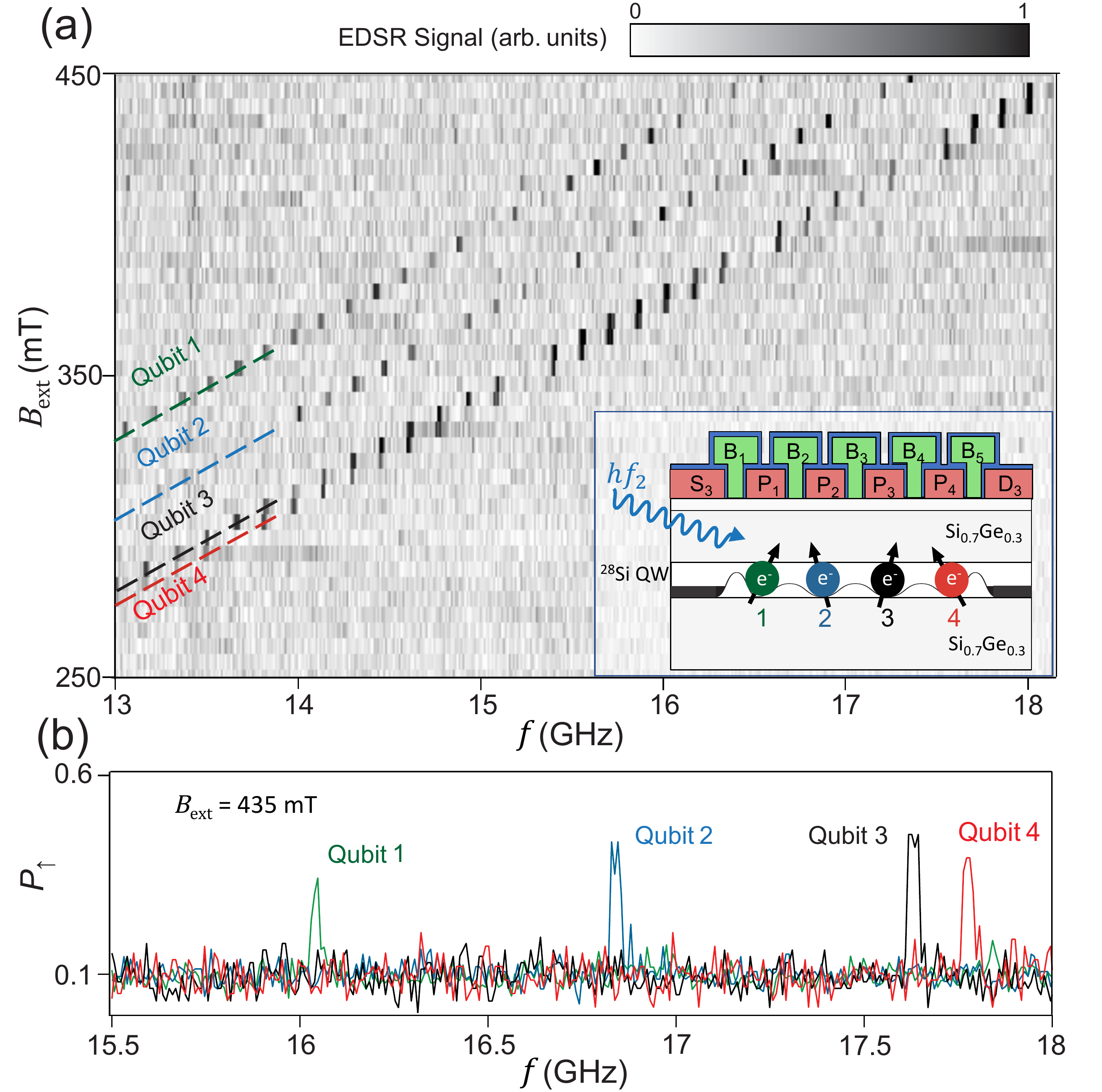}
		\caption{
			(a) Summed EDSR spectra for all four qubits. The qubits are driven using gate MS and four distinct ESR transitions are observed. Inset: Cross-sectional cartoon of the device showing all four qubits subject to the same driving field, but with only qubit 2 on resonance. (b) A line cut through the data at $B_{ext}$ = 435 mT shows that all four qubits are well separated in frequency. The spacing between qubits 1, 2, and 3 is $\sim$800 MHz and qubits 3 and 4 are separated by $\sim$130 MHz. The qubit linewidths are broadened by the 30 MHz microwave chirp used in these measurements.
		}
		\label{Fig3}
	\end{center}
	\vspace{-0.6cm}
\end{figure}

\begin{figure*}[t]
	\begin{center}
		\includegraphics[width=\textwidth]{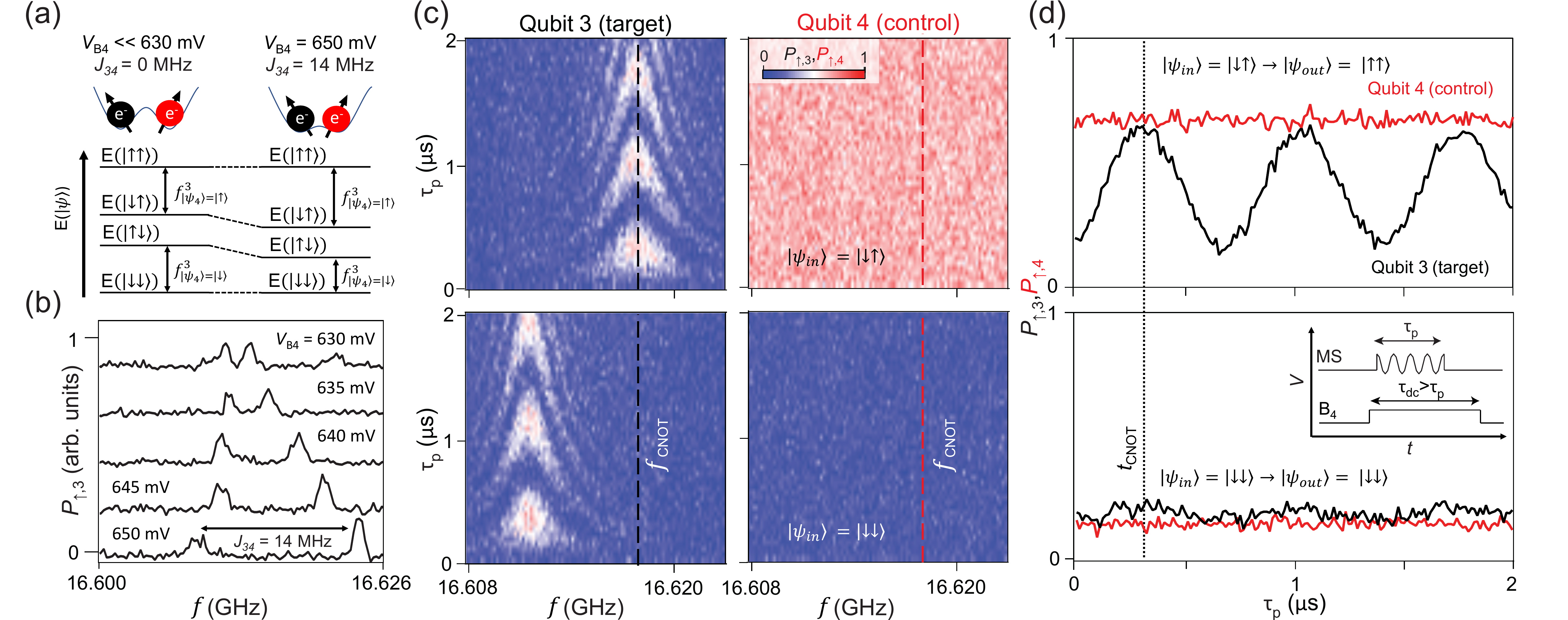}
		\caption{(a) Energy level diagram for dots 3 and 4 subject to a magnetic field gradient and either no exchange (left) or finite exchange (right). Increasing $V_{B4}$ turns on the exchange interaction $J_{34}$ between qubits 3 and 4. Under exchange, the spin transition frequency of qubit 3 (the target qubit) $f^{(3)}$ depends on the input state of qubit 4 $\ket{\Psi_{4}}$ (the control qubit). (b) EDSR spectroscopy of qubit 3. The spin-up probability $P_{\uparrow,3}$ is plotted as a function of $f$ and $V_{B4}$. (c) By varying the frequency and amplitude of a microwave pulse applied to the target qubit with the control qubit prepared in either the up state (top panels) or the down state (bottom panels), we map out the frequency of the target qubit conditioned on the state of the control qubit. (d) Driving at frequency $f_{\text{CNOT}}$ and varying the microwave burst time $\tau_{p}$ results in Rabi oscillations of the target qubit that are conditioned on the state of the control qubit. When the microwave burst is timed to correspond to a $\pi$-rotation on the target qubit, $\tau_{p}$ = $\tau_{\text{CNOT}}=$ 270 ns, a CNOT gate is achieved (dotted line).}
		\label{Fig4}
	\end{center}
	\vspace{-0.6cm}
\end{figure*}

Two-qubit gates are performed by applying voltage pulses to the barrier gates $B_{i}$ separating the dots \cite{loss1998}. Applying a positive voltage to a barrier gate increases wavefunction overlap between adjacent dots and turns on exchange \cite{Russ2018}, as illustrated in the energy level diagram of Fig.~4(a). To first map out the influence of the barrier gate on the exchange interaction, we vary the gate voltage $V_{B4}$ and perform EDSR spectroscopy on qubit 3, as shown in Fig.~4(b). Before measuring the EDSR spectrum of qubit 3, qubit 4 was prepared in a mixed state such that any state dependent line splitting could be observed \cite{Zajac2018}. The EDSR spectrum of qubit 4 is probed in a similar way. As $V_{B4}$ is increased, the EDSR lines split to reveal a doublet with a peak separation that is equal to the exchange energy $J_{34}$. The overall frequency shift of the doublet is attributed to the displacement of the electron's wavefunction within the magnetic field gradient as $V_{B4}$ is adjusted \cite{Zajac2018}.

A resonant-CNOT gate can be achieved using this device architecture by following the protocol developed by Zajac \textit{et al.} \cite{Zajac2018}. Here we focus on qubits 3 and 4. With $J_{34}\approx0$, we prepare input state $\ket{\psi_{in}}$ = $\ket{\downarrow\downarrow}$ through spin-selective tunneling or input state $\ket{\psi_{in}}$ = $\ket{\downarrow\uparrow}$ through spin-selective tunneling followed by a $\pi$-pulse on spin 4. We then apply a voltage pulse to gate $B_{4}$ to turn on exchange while simultaneously applying a microwave burst of varying duration $\tau_{p}$ and frequency $f$. The resulting spin-up probabilities $P_{\uparrow,3}$($P_{\uparrow,4}$) for qubit 3(4) are plotted in Fig.~4(c). Rabi oscillations are observed for qubit 3 with a resonance frequency that is dependent on the state of qubit 4. By setting $f$ = $f_{\text{CNOT}}$, where the microwave tone is resonant with the target qubit when the control qubit is in the spin-up state, we drive Rabi oscillations on the target qubit conditioned on the state of the control qubit [see Fig.~4(d)]. Furthermore, when $t$ = $t_{\text{CNOT}}$, qubit 3 will be flipped when qubit 4 is in the spin-up state. With these settings, we realize a resonant-CNOT gate in a four qubit device \cite{Zajac2018,Russ2018}. To implement a high fidelity CNOT gate, it is important to note that some additional phase accumulation due to the $V_{\text{B4}}$-induced Ising interaction must be compensated for by properly tuning $\tau_{\text{DC}}$ and $V_{\text{B4}}$ as described in our previous work \cite{Zajac2018,Russ2018}. We attribute the slight oscillations that appear on qubit 3 in the lower panel of Fig.~4(d) to state preparation errors on qubit 4. Off-resonant driving would lead to oscillations at a frequency above 6 MHz.

In conclusion, we have demonstrated one- and two-qubit gate operations in a four qubit device fabricated from an isotopically enriched $^{28}$Si quantum well. Our device design allows for full control of the charge state in the array. Interdot tunnel coupling and exchange are tuned using barrier gates. We demonstrate independent control of all four qubits, which is enabled by the field gradient from a Co micromagnet. To demonstrate a two-qubit gate involving dots 3 and 4, we mapped exchange as a function of $V_{B4}$ and performed a resonant-CNOT gate in 270 ns. These results set the stage for a four qubit spin-based quantum processor in silicon, which should be capable of performing small-scale quantum algorithms and demonstrating time crystal phases \cite{Barnestc}.

Funded by Army Research Office grant No.\ W911NF-15-1-0149, DARPA grant No.\ D18AC0025, and the Gordon and Betty Moore Foundation's EPiQS Initiative through Grant GBMF4535. Devices were fabricated in the Princeton University Quantum Device Nanofabrication Laboratory.


\begin{thebibliography}{45}%
\makeatletter
\providecommand \@ifxundefined [1]{%
 \@ifx{#1\undefined}
}%
\providecommand \@ifnum [1]{%
 \ifnum #1\expandafter \@firstoftwo
 \else \expandafter \@secondoftwo
 \fi
}%
\providecommand \@ifx [1]{%
 \ifx #1\expandafter \@firstoftwo
 \else \expandafter \@secondoftwo
 \fi
}%
\providecommand \natexlab [1]{#1}%
\providecommand \enquote  [1]{``#1''}%
\providecommand \bibnamefont  [1]{#1}%
\providecommand \bibfnamefont [1]{#1}%
\providecommand \citenamefont [1]{#1}%
\providecommand \href@noop [0]{\@secondoftwo}%
\providecommand \href [0]{\begingroup \@sanitize@url \@href}%
\providecommand \@href[1]{\@@startlink{#1}\@@href}%
\providecommand \@@href[1]{\endgroup#1\@@endlink}%
\providecommand \@sanitize@url [0]{\catcode `\\12\catcode `\$12\catcode
  `\&12\catcode `\#12\catcode `\^12\catcode `\_12\catcode `\%12\relax}%
\providecommand \@@startlink[1]{}%
\providecommand \@@endlink[0]{}%
\providecommand \url  [0]{\begingroup\@sanitize@url \@url }%
\providecommand \@url [1]{\endgroup\@href {#1}{\urlprefix }}%
\providecommand \urlprefix  [0]{URL }%
\providecommand \Eprint [0]{\href }%
\providecommand \doibase [0]{http://dx.doi.org/}%
\providecommand \selectlanguage [0]{\@gobble}%
\providecommand \bibinfo  [0]{\@secondoftwo}%
\providecommand \bibfield  [0]{\@secondoftwo}%
\providecommand \translation [1]{[#1]}%
\providecommand \BibitemOpen [0]{}%
\providecommand \bibitemStop [0]{}%
\providecommand \bibitemNoStop [0]{.\EOS\space}%
\providecommand \EOS [0]{\spacefactor3000\relax}%
\providecommand \BibitemShut  [1]{\csname bibitem#1\endcsname}%
\let\auto@bib@innerbib\@empty
\bibitem [{\citenamefont {Loss}\ and\ \citenamefont
  {DiVincenzo}(1998)}]{loss1998}%
  \BibitemOpen
  \bibfield  {author} {\bibinfo {author} {\bibfnamefont {D.}~\bibnamefont
  {Loss}}\ and\ \bibinfo {author} {\bibfnamefont {D.~P.}\ \bibnamefont
  {DiVincenzo}},\ }\bibfield  {title} {\enquote {\bibinfo {title} {Quantum
  computation with quantum dots},}\ }\href {\doibase 10.1103/PhysRevA.57.120}
  {\bibfield  {journal} {\bibinfo  {journal} {Phys. Rev. A}\ }\textbf {\bibinfo
  {volume} {57}},\ \bibinfo {pages} {120} (\bibinfo {year} {1998})}\BibitemShut
  {NoStop}%
\bibitem [{\citenamefont {Hanson}\ \emph {et~al.}(2007)\citenamefont {Hanson},
  \citenamefont {Kouwenhoven}, \citenamefont {Petta}, \citenamefont {Tarucha},\
  and\ \citenamefont {Vandersypen}}]{hanson2007spins}%
  \BibitemOpen
  \bibfield  {author} {\bibinfo {author} {\bibfnamefont {R.}~\bibnamefont
  {Hanson}}, \bibinfo {author} {\bibfnamefont {L.~P.}\ \bibnamefont
  {Kouwenhoven}}, \bibinfo {author} {\bibfnamefont {J.~R.}\ \bibnamefont
  {Petta}}, \bibinfo {author} {\bibfnamefont {S.}~\bibnamefont {Tarucha}}, \
  and\ \bibinfo {author} {\bibfnamefont {L.~M.~K.}\ \bibnamefont
  {Vandersypen}},\ }\bibfield  {title} {\enquote {\bibinfo {title} {Spins in
  few-electron quantum dots},}\ }\href@noop {} {\bibfield  {journal} {\bibinfo
  {journal} {Rev. Mod. Phys.}\ }\textbf {\bibinfo {volume} {79}},\ \bibinfo
  {pages} {1217} (\bibinfo {year} {2007})}\BibitemShut {NoStop}%
\bibitem [{\citenamefont {Zwanenburg}\ \emph {et~al.}(2013)\citenamefont
  {Zwanenburg}, \citenamefont {Dzurak}, \citenamefont {Morello}, \citenamefont
  {Simmons}, \citenamefont {Hollenberg}, \citenamefont {Klimeck}, \citenamefont
  {Rogge}, \citenamefont {Coppersmith},\ and\ \citenamefont
  {Eriksson}}]{zwanenburg2013silicon}%
  \BibitemOpen
  \bibfield  {author} {\bibinfo {author} {\bibfnamefont {F.~A.}\ \bibnamefont
  {Zwanenburg}}, \bibinfo {author} {\bibfnamefont {A.~S.}\ \bibnamefont
  {Dzurak}}, \bibinfo {author} {\bibfnamefont {A.}~\bibnamefont {Morello}},
  \bibinfo {author} {\bibfnamefont {M.~Y.}\ \bibnamefont {Simmons}}, \bibinfo
  {author} {\bibfnamefont {L.~C.~L.}\ \bibnamefont {Hollenberg}}, \bibinfo
  {author} {\bibfnamefont {G.}~\bibnamefont {Klimeck}}, \bibinfo {author}
  {\bibfnamefont {S.}~\bibnamefont {Rogge}}, \bibinfo {author} {\bibfnamefont
  {S.~N.}\ \bibnamefont {Coppersmith}}, \ and\ \bibinfo {author} {\bibfnamefont
  {M.~A.}\ \bibnamefont {Eriksson}},\ }\bibfield  {title} {\enquote {\bibinfo
  {title} {Silicon quantum electronics},}\ }\href@noop {} {\bibfield  {journal}
  {\bibinfo  {journal} {Rev. Mod. Phys.}\ }\textbf {\bibinfo {volume} {85}},\
  \bibinfo {pages} {961} (\bibinfo {year} {2013})}\BibitemShut {NoStop}%
\bibitem [{\citenamefont {Tyryshkin}\ \emph {et~al.}(2012)\citenamefont
  {Tyryshkin}, \citenamefont {Tojo}, \citenamefont {Morton}, \citenamefont
  {Riemann}, \citenamefont {Abrosimov}, \citenamefont {Becker}, \citenamefont
  {Pohl}, \citenamefont {Schenkel}, \citenamefont {Thewalt}, \citenamefont
  {Itoh},\ and\ \citenamefont {Lyon}}]{tyryshkin2012electron}%
  \BibitemOpen
  \bibfield  {author} {\bibinfo {author} {\bibfnamefont {A.~M.}\ \bibnamefont
  {Tyryshkin}}, \bibinfo {author} {\bibfnamefont {S.}~\bibnamefont {Tojo}},
  \bibinfo {author} {\bibfnamefont {J.~J.~L.}\ \bibnamefont {Morton}}, \bibinfo
  {author} {\bibfnamefont {H.}~\bibnamefont {Riemann}}, \bibinfo {author}
  {\bibfnamefont {N.~V.}\ \bibnamefont {Abrosimov}}, \bibinfo {author}
  {\bibfnamefont {P.}~\bibnamefont {Becker}}, \bibinfo {author} {\bibfnamefont
  {H.-J.}\ \bibnamefont {Pohl}}, \bibinfo {author} {\bibfnamefont
  {T.}~\bibnamefont {Schenkel}}, \bibinfo {author} {\bibfnamefont {M.~L.~W.}\
  \bibnamefont {Thewalt}}, \bibinfo {author} {\bibfnamefont {K.~M.}\
  \bibnamefont {Itoh}}, \ and\ \bibinfo {author} {\bibfnamefont {S.~A.}\
  \bibnamefont {Lyon}},\ }\bibfield  {title} {\enquote {\bibinfo {title}
  {Electron spin coherence exceeding seconds in high-purity silicon},}\
  }\href@noop {} {\bibfield  {journal} {\bibinfo  {journal} {Nat. Mater.}\
  }\textbf {\bibinfo {volume} {11}},\ \bibinfo {pages} {143} (\bibinfo {year}
  {2012})}\BibitemShut {NoStop}%
\bibitem [{\citenamefont {Veldhorst}\ \emph {et~al.}(2014)\citenamefont
  {Veldhorst}, \citenamefont {Hwang}, \citenamefont {Yang}, \citenamefont
  {Leenstra}, \citenamefont {De~Ronde}, \citenamefont {Dehollain},
  \citenamefont {Muhonen}, \citenamefont {Hudson}, \citenamefont {Itoh},
  \citenamefont {Morello},\ and\ \citenamefont
  {Dzurak}}]{veldhorst2014addressable}%
  \BibitemOpen
  \bibfield  {author} {\bibinfo {author} {\bibfnamefont {M.}~\bibnamefont
  {Veldhorst}}, \bibinfo {author} {\bibfnamefont {J.~C.~C.}\ \bibnamefont
  {Hwang}}, \bibinfo {author} {\bibfnamefont {C.~H.}\ \bibnamefont {Yang}},
  \bibinfo {author} {\bibfnamefont {A.~W.}\ \bibnamefont {Leenstra}}, \bibinfo
  {author} {\bibfnamefont {B.}~\bibnamefont {De~Ronde}}, \bibinfo {author}
  {\bibfnamefont {J.~P.}\ \bibnamefont {Dehollain}}, \bibinfo {author}
  {\bibfnamefont {J.~T.}\ \bibnamefont {Muhonen}}, \bibinfo {author}
  {\bibfnamefont {F.~E.}\ \bibnamefont {Hudson}}, \bibinfo {author}
  {\bibfnamefont {K.~M.}\ \bibnamefont {Itoh}}, \bibinfo {author}
  {\bibfnamefont {A.}~\bibnamefont {Morello}}, \ and\ \bibinfo {author}
  {\bibfnamefont {A.~S.}\ \bibnamefont {Dzurak}},\ }\bibfield  {title}
  {\enquote {\bibinfo {title} {An addressable quantum dot qubit with
  fault-tolerant control-fidelity},}\ }\href@noop {} {\bibfield  {journal}
  {\bibinfo  {journal} {Nat. Nanotechnol.}\ }\textbf {\bibinfo {volume} {9}},\
  \bibinfo {pages} {981} (\bibinfo {year} {2014})}\BibitemShut {NoStop}%
\bibitem [{\citenamefont {Yoneda}\ \emph {et~al.}(2018)\citenamefont {Yoneda},
  \citenamefont {Takeda}, \citenamefont {Otsuka}, \citenamefont {Nakajima},
  \citenamefont {Delbecq}, \citenamefont {Allison}, \citenamefont {Honda},
  \citenamefont {Kodera}, \citenamefont {Oda}, \citenamefont {Hoshi},
  \citenamefont {Usami}, \citenamefont {Itoh},\ and\ \citenamefont
  {Tarucha}}]{Yoneda2018}%
  \BibitemOpen
  \bibfield  {author} {\bibinfo {author} {\bibfnamefont {J.}~\bibnamefont
  {Yoneda}}, \bibinfo {author} {\bibfnamefont {K.}~\bibnamefont {Takeda}},
  \bibinfo {author} {\bibfnamefont {T.}~\bibnamefont {Otsuka}}, \bibinfo
  {author} {\bibfnamefont {T.}~\bibnamefont {Nakajima}}, \bibinfo {author}
  {\bibfnamefont {M.~R.}\ \bibnamefont {Delbecq}}, \bibinfo {author}
  {\bibfnamefont {G.}~\bibnamefont {Allison}}, \bibinfo {author} {\bibfnamefont
  {T.}~\bibnamefont {Honda}}, \bibinfo {author} {\bibfnamefont
  {T.}~\bibnamefont {Kodera}}, \bibinfo {author} {\bibfnamefont
  {S.}~\bibnamefont {Oda}}, \bibinfo {author} {\bibfnamefont {Y.}~\bibnamefont
  {Hoshi}}, \bibinfo {author} {\bibfnamefont {N.}~\bibnamefont {Usami}},
  \bibinfo {author} {\bibfnamefont {K.~M.}\ \bibnamefont {Itoh}}, \ and\
  \bibinfo {author} {\bibfnamefont {S.}~\bibnamefont {Tarucha}},\ }\bibfield
  {title} {\enquote {\bibinfo {title} {A quantum-dot spin qubit with coherence
  limited by charge noise and fidelity higher than 99.9\%},}\ }\href {\doibase
  10.1038/s41565-017-0014-x} {\bibfield  {journal} {\bibinfo  {journal} {Nat.
  Nanotechnol.}\ }\textbf {\bibinfo {volume} {13}},\ \bibinfo {pages} {102}
  (\bibinfo {year} {2018})}\BibitemShut {NoStop}%
\bibitem [{\citenamefont {{Yang}}\ \emph {et~al.}(2018)\citenamefont {{Yang}},
  \citenamefont {{Chan}}, \citenamefont {{Harper}}, \citenamefont {{Huang}},
  \citenamefont {{Evans}}, \citenamefont {{Hwang}}, \citenamefont {{Hensen}},
  \citenamefont {{Laucht}}, \citenamefont {{Tanttu}}, \citenamefont {{Hudson}},
  \citenamefont {{Flammia}}, \citenamefont {{Itoh}}, \citenamefont {{Morello}},
  \citenamefont {{Bartlett}},\ and\ \citenamefont {{Dzurak}}}]{Yang2018}%
  \BibitemOpen
  \bibfield  {author} {\bibinfo {author} {\bibfnamefont {C.~H.}\ \bibnamefont
  {{Yang}}}, \bibinfo {author} {\bibfnamefont {K.~W.}\ \bibnamefont {{Chan}}},
  \bibinfo {author} {\bibfnamefont {R.}~\bibnamefont {{Harper}}}, \bibinfo
  {author} {\bibfnamefont {W.}~\bibnamefont {{Huang}}}, \bibinfo {author}
  {\bibfnamefont {T.}~\bibnamefont {{Evans}}}, \bibinfo {author} {\bibfnamefont
  {J.~C.~C.}\ \bibnamefont {{Hwang}}}, \bibinfo {author} {\bibfnamefont
  {B.}~\bibnamefont {{Hensen}}}, \bibinfo {author} {\bibfnamefont
  {A.}~\bibnamefont {{Laucht}}}, \bibinfo {author} {\bibfnamefont
  {T.}~\bibnamefont {{Tanttu}}}, \bibinfo {author} {\bibfnamefont {F.~E.}\
  \bibnamefont {{Hudson}}}, \bibinfo {author} {\bibfnamefont {S.~T.}\
  \bibnamefont {{Flammia}}}, \bibinfo {author} {\bibfnamefont {K.~M.}\
  \bibnamefont {{Itoh}}}, \bibinfo {author} {\bibfnamefont {A.}~\bibnamefont
  {{Morello}}}, \bibinfo {author} {\bibfnamefont {S.~D.}\ \bibnamefont
  {{Bartlett}}}, \ and\ \bibinfo {author} {\bibfnamefont {A.~S.}\ \bibnamefont
  {{Dzurak}}},\ }\bibfield  {title} {\enquote {\bibinfo {title} {{Silicon qubit
  fidelities approaching incoherent noise limits via pulse optimisation}},}\
  }\href@noop {} {\bibfield  {journal} {\bibinfo  {journal} {arXiv:1807.09500}\
  } (\bibinfo {year} {2018})}\BibitemShut {NoStop}%
\bibitem [{\citenamefont {Veldhorst}\ \emph {et~al.}(2015)\citenamefont
  {Veldhorst}, \citenamefont {Yang}, \citenamefont {Hwang}, \citenamefont
  {Huang}, \citenamefont {Dehollain}, \citenamefont {Muhonen}, \citenamefont
  {Simmons}, \citenamefont {Laucht}, \citenamefont {Hudson}, \citenamefont
  {Itoh}, \citenamefont {Morello},\ and\ \citenamefont
  {Dzurak}}]{Veldhorst2015}%
  \BibitemOpen
  \bibfield  {author} {\bibinfo {author} {\bibfnamefont {M.}~\bibnamefont
  {Veldhorst}}, \bibinfo {author} {\bibfnamefont {C.~H.}\ \bibnamefont {Yang}},
  \bibinfo {author} {\bibfnamefont {J.~C.~C.}\ \bibnamefont {Hwang}}, \bibinfo
  {author} {\bibfnamefont {W.}~\bibnamefont {Huang}}, \bibinfo {author}
  {\bibfnamefont {J.~P.}\ \bibnamefont {Dehollain}}, \bibinfo {author}
  {\bibfnamefont {J.~T.}\ \bibnamefont {Muhonen}}, \bibinfo {author}
  {\bibfnamefont {S.}~\bibnamefont {Simmons}}, \bibinfo {author} {\bibfnamefont
  {A.}~\bibnamefont {Laucht}}, \bibinfo {author} {\bibfnamefont {F.~E.}\
  \bibnamefont {Hudson}}, \bibinfo {author} {\bibfnamefont {K.~M.}\
  \bibnamefont {Itoh}}, \bibinfo {author} {\bibfnamefont {A.}~\bibnamefont
  {Morello}}, \ and\ \bibinfo {author} {\bibfnamefont {A.~S.}\ \bibnamefont
  {Dzurak}},\ }\bibfield  {title} {\enquote {\bibinfo {title} {A two-qubit
  logic gate in silicon},}\ }\href {https://doi.org/10.1038/nature15263}
  {\bibfield  {journal} {\bibinfo  {journal} {Nature}\ }\textbf {\bibinfo
  {volume} {526}},\ \bibinfo {pages} {410} (\bibinfo {year}
  {2015})}\BibitemShut {NoStop}%
\bibitem [{\citenamefont {Zajac}\ \emph {et~al.}(2018)\citenamefont {Zajac},
  \citenamefont {Sigillito}, \citenamefont {Russ}, \citenamefont {Borjans},
  \citenamefont {Taylor}, \citenamefont {Burkard},\ and\ \citenamefont
  {Petta}}]{Zajac2018}%
  \BibitemOpen
  \bibfield  {author} {\bibinfo {author} {\bibfnamefont {D.~M.}\ \bibnamefont
  {Zajac}}, \bibinfo {author} {\bibfnamefont {A.~J.}\ \bibnamefont
  {Sigillito}}, \bibinfo {author} {\bibfnamefont {M.}~\bibnamefont {Russ}},
  \bibinfo {author} {\bibfnamefont {F.}~\bibnamefont {Borjans}}, \bibinfo
  {author} {\bibfnamefont {J.~M.}\ \bibnamefont {Taylor}}, \bibinfo {author}
  {\bibfnamefont {G.}~\bibnamefont {Burkard}}, \ and\ \bibinfo {author}
  {\bibfnamefont {J.~R.}\ \bibnamefont {Petta}},\ }\bibfield  {title} {\enquote
  {\bibinfo {title} {Resonantly driven {CNOT} gate for electron spins},}\
  }\href {\doibase 10.1126/science.aao5965} {\bibfield  {journal} {\bibinfo
  {journal} {Science}\ }\textbf {\bibinfo {volume} {359}},\ \bibinfo {pages}
  {439} (\bibinfo {year} {2018})}\BibitemShut {NoStop}%
\bibitem [{\citenamefont {Watson}\ \emph {et~al.}(2018)\citenamefont {Watson},
  \citenamefont {Philips}, \citenamefont {Kawakami}, \citenamefont {Ward},
  \citenamefont {Scarlino}, \citenamefont {Veldhorst}, \citenamefont {Savage},
  \citenamefont {Lagally}, \citenamefont {Friesen}, \citenamefont
  {Coppersmith}, \citenamefont {Eriksson},\ and\ \citenamefont
  {Vandersypen}}]{Watson2018}%
  \BibitemOpen
  \bibfield  {author} {\bibinfo {author} {\bibfnamefont {T.~F.}\ \bibnamefont
  {Watson}}, \bibinfo {author} {\bibfnamefont {S.~G.~J.}\ \bibnamefont
  {Philips}}, \bibinfo {author} {\bibfnamefont {E.}~\bibnamefont {Kawakami}},
  \bibinfo {author} {\bibfnamefont {D.~R.}\ \bibnamefont {Ward}}, \bibinfo
  {author} {\bibfnamefont {P.}~\bibnamefont {Scarlino}}, \bibinfo {author}
  {\bibfnamefont {M.}~\bibnamefont {Veldhorst}}, \bibinfo {author}
  {\bibfnamefont {D.~E.}\ \bibnamefont {Savage}}, \bibinfo {author}
  {\bibfnamefont {M.~G.}\ \bibnamefont {Lagally}}, \bibinfo {author}
  {\bibfnamefont {M.}~\bibnamefont {Friesen}}, \bibinfo {author} {\bibfnamefont
  {S.~N.}\ \bibnamefont {Coppersmith}}, \bibinfo {author} {\bibfnamefont
  {M.~A.}\ \bibnamefont {Eriksson}}, \ and\ \bibinfo {author} {\bibfnamefont
  {L.~M.~K.}\ \bibnamefont {Vandersypen}},\ }\bibfield  {title} {\enquote
  {\bibinfo {title} {A programmable two-qubit quantum processor in silicon},}\
  }\href {https://doi.org/10.1038/nature25766} {\bibfield  {journal} {\bibinfo
  {journal} {Nature}\ }\textbf {\bibinfo {volume} {555}},\ \bibinfo {pages}
  {633} (\bibinfo {year} {2018})}\BibitemShut {NoStop}%
\bibitem [{\citenamefont {{Huang}}\ \emph {et~al.}(2018)\citenamefont
  {{Huang}}, \citenamefont {{Yang}}, \citenamefont {{Chan}}, \citenamefont
  {{Tanttu}}, \citenamefont {{Hensen}}, \citenamefont {{Leon}}, \citenamefont
  {{Fogarty}}, \citenamefont {{Hwang}}, \citenamefont {{Hudson}}, \citenamefont
  {{Itoh}}, \citenamefont {{Morello}}, \citenamefont {{Laucht}},\ and\
  \citenamefont {{Dzurak}}}]{Huang2018}%
  \BibitemOpen
  \bibfield  {author} {\bibinfo {author} {\bibfnamefont {W.}~\bibnamefont
  {{Huang}}}, \bibinfo {author} {\bibfnamefont {C.~H.}\ \bibnamefont {{Yang}}},
  \bibinfo {author} {\bibfnamefont {K.~W.}\ \bibnamefont {{Chan}}}, \bibinfo
  {author} {\bibfnamefont {T.}~\bibnamefont {{Tanttu}}}, \bibinfo {author}
  {\bibfnamefont {B.}~\bibnamefont {{Hensen}}}, \bibinfo {author}
  {\bibfnamefont {R.~C.~C.}\ \bibnamefont {{Leon}}}, \bibinfo {author}
  {\bibfnamefont {M.~A.}\ \bibnamefont {{Fogarty}}}, \bibinfo {author}
  {\bibfnamefont {J.~C.~C.}\ \bibnamefont {{Hwang}}}, \bibinfo {author}
  {\bibfnamefont {F.~E.}\ \bibnamefont {{Hudson}}}, \bibinfo {author}
  {\bibfnamefont {K.~M.}\ \bibnamefont {{Itoh}}}, \bibinfo {author}
  {\bibfnamefont {A.}~\bibnamefont {{Morello}}}, \bibinfo {author}
  {\bibfnamefont {A.}~\bibnamefont {{Laucht}}}, \ and\ \bibinfo {author}
  {\bibfnamefont {A.~S.}\ \bibnamefont {{Dzurak}}},\ }\bibfield  {title}
  {\enquote {\bibinfo {title} {{Fidelity benchmarks for two-qubit gates in
  silicon}},}\ }\href@noop {} {\bibfield  {journal} {\bibinfo  {journal}
  {arXiv:1805.05027}\ } (\bibinfo {year} {2018})}\BibitemShut {NoStop}%
\bibitem [{\citenamefont {{Xue}}\ \emph {et~al.}(2018)\citenamefont {{Xue}},
  \citenamefont {{Watson}}, \citenamefont {{Helsen}}, \citenamefont {{Ward}},
  \citenamefont {{Savage}}, \citenamefont {{Lagally}}, \citenamefont
  {{Coppersmith}}, \citenamefont {{Eriksson}}, \citenamefont {{Wehner}},\ and\
  \citenamefont {{Vandersypen}}}]{Xue2018}%
  \BibitemOpen
  \bibfield  {author} {\bibinfo {author} {\bibfnamefont {X.}~\bibnamefont
  {{Xue}}}, \bibinfo {author} {\bibfnamefont {T.~F.}\ \bibnamefont {{Watson}}},
  \bibinfo {author} {\bibfnamefont {J.}~\bibnamefont {{Helsen}}}, \bibinfo
  {author} {\bibfnamefont {D.~R.}\ \bibnamefont {{Ward}}}, \bibinfo {author}
  {\bibfnamefont {D.~E.}\ \bibnamefont {{Savage}}}, \bibinfo {author}
  {\bibfnamefont {M.~G.}\ \bibnamefont {{Lagally}}}, \bibinfo {author}
  {\bibfnamefont {S.~N.}\ \bibnamefont {{Coppersmith}}}, \bibinfo {author}
  {\bibfnamefont {M.~A.}\ \bibnamefont {{Eriksson}}}, \bibinfo {author}
  {\bibfnamefont {S.}~\bibnamefont {{Wehner}}}, \ and\ \bibinfo {author}
  {\bibfnamefont {L.~M.~K.}\ \bibnamefont {{Vandersypen}}},\ }\bibfield
  {title} {\enquote {\bibinfo {title} {{Benchmarking Gate Fidelities in a
  Si/SiGe Two-Qubit Device}},}\ }\href@noop {} {\bibfield  {journal} {\bibinfo
  {journal} {arXiv:1811.04002}\ } (\bibinfo {year} {2018})}\BibitemShut
  {NoStop}%
\bibitem [{\citenamefont {Zajac}\ \emph {et~al.}(2016)\citenamefont {Zajac},
  \citenamefont {Hazard}, \citenamefont {Mi}, \citenamefont {Nielsen},\ and\
  \citenamefont {Petta}}]{ZajacScalable}%
  \BibitemOpen
  \bibfield  {author} {\bibinfo {author} {\bibfnamefont {D.~M.}\ \bibnamefont
  {Zajac}}, \bibinfo {author} {\bibfnamefont {T.~M.}\ \bibnamefont {Hazard}},
  \bibinfo {author} {\bibfnamefont {X.}~\bibnamefont {Mi}}, \bibinfo {author}
  {\bibfnamefont {E.}~\bibnamefont {Nielsen}}, \ and\ \bibinfo {author}
  {\bibfnamefont {J.~R.}\ \bibnamefont {Petta}},\ }\bibfield  {title} {\enquote
  {\bibinfo {title} {Scalable gate architecture for a one-dimensional array of
  semiconductor spin qubits},}\ }\href {\doibase
  10.1103/PhysRevApplied.6.054013} {\bibfield  {journal} {\bibinfo  {journal}
  {Phys. Rev. Appl.}\ }\textbf {\bibinfo {volume} {6}},\ \bibinfo {pages}
  {054013} (\bibinfo {year} {2016})}\BibitemShut {NoStop}%
\bibitem [{\citenamefont {Noiri}\ \emph {et~al.}(2016)\citenamefont {Noiri},
  \citenamefont {Yoneda}, \citenamefont {Nakajima}, \citenamefont {Otsuka},
  \citenamefont {Delbecq}, \citenamefont {Takeda}, \citenamefont {Amaha},
  \citenamefont {Allison}, \citenamefont {Ludwig}, \citenamefont {Wieck},\ and\
  \citenamefont {Tarucha}}]{Noiri2016}%
  \BibitemOpen
  \bibfield  {author} {\bibinfo {author} {\bibfnamefont {A.}~\bibnamefont
  {Noiri}}, \bibinfo {author} {\bibfnamefont {J.}~\bibnamefont {Yoneda}},
  \bibinfo {author} {\bibfnamefont {T.}~\bibnamefont {Nakajima}}, \bibinfo
  {author} {\bibfnamefont {T.}~\bibnamefont {Otsuka}}, \bibinfo {author}
  {\bibfnamefont {M.~R.}\ \bibnamefont {Delbecq}}, \bibinfo {author}
  {\bibfnamefont {K.}~\bibnamefont {Takeda}}, \bibinfo {author} {\bibfnamefont
  {S.}~\bibnamefont {Amaha}}, \bibinfo {author} {\bibfnamefont
  {G.}~\bibnamefont {Allison}}, \bibinfo {author} {\bibfnamefont
  {A.}~\bibnamefont {Ludwig}}, \bibinfo {author} {\bibfnamefont {A.~D.}\
  \bibnamefont {Wieck}}, \ and\ \bibinfo {author} {\bibfnamefont
  {S.}~\bibnamefont {Tarucha}},\ }\bibfield  {title} {\enquote {\bibinfo
  {title} {Coherent electron-spin-resonance manipulation of three individual
  spins in a triple quantum dot},}\ }\href {\doibase 10.1063/1.4945592}
  {\bibfield  {journal} {\bibinfo  {journal} {Appl. Phys. Lett.}\ }\textbf
  {\bibinfo {volume} {108}},\ \bibinfo {pages} {153101} (\bibinfo {year}
  {2016})}\BibitemShut {NoStop}%
\bibitem [{\citenamefont {Otsuka}\ \emph {et~al.}(2016)\citenamefont {Otsuka},
  \citenamefont {Nakajima}, \citenamefont {Delbecq}, \citenamefont {Amaha},
  \citenamefont {Yoneda}, \citenamefont {Takeda}, \citenamefont {Allison},
  \citenamefont {Ito}, \citenamefont {Sugawara}, \citenamefont {Noiri},
  \citenamefont {Ludwig}, \citenamefont {Wieck},\ and\ \citenamefont
  {Tarucha}}]{Otsuka2016}%
  \BibitemOpen
  \bibfield  {author} {\bibinfo {author} {\bibfnamefont {T.}~\bibnamefont
  {Otsuka}}, \bibinfo {author} {\bibfnamefont {T.}~\bibnamefont {Nakajima}},
  \bibinfo {author} {\bibfnamefont {M.~R.}\ \bibnamefont {Delbecq}}, \bibinfo
  {author} {\bibfnamefont {S.}~\bibnamefont {Amaha}}, \bibinfo {author}
  {\bibfnamefont {J.}~\bibnamefont {Yoneda}}, \bibinfo {author} {\bibfnamefont
  {K.}~\bibnamefont {Takeda}}, \bibinfo {author} {\bibfnamefont
  {G.}~\bibnamefont {Allison}}, \bibinfo {author} {\bibfnamefont
  {T.}~\bibnamefont {Ito}}, \bibinfo {author} {\bibfnamefont {R.}~\bibnamefont
  {Sugawara}}, \bibinfo {author} {\bibfnamefont {A.}~\bibnamefont {Noiri}},
  \bibinfo {author} {\bibfnamefont {A.}~\bibnamefont {Ludwig}}, \bibinfo
  {author} {\bibfnamefont {A.~D.}\ \bibnamefont {Wieck}}, \ and\ \bibinfo
  {author} {\bibfnamefont {S.}~\bibnamefont {Tarucha}},\ }\bibfield  {title}
  {\enquote {\bibinfo {title} {Single-electron spin resonance in a quadruple
  quantum dot},}\ }\href@noop {} {\bibfield  {journal} {\bibinfo  {journal}
  {Sci. Rep.}\ }\textbf {\bibinfo {volume} {6}},\ \bibinfo {pages} {31820}
  (\bibinfo {year} {2016})}\BibitemShut {NoStop}%
\bibitem [{\citenamefont {Ito}\ \emph {et~al.}(2018)\citenamefont {Ito},
  \citenamefont {Otsuka}, \citenamefont {Nakajima}, \citenamefont {Delbecq},
  \citenamefont {Amaha}, \citenamefont {Yoneda}, \citenamefont {Takeda},
  \citenamefont {Noiri}, \citenamefont {Allison}, \citenamefont {Ludwig},
  \citenamefont {Wieck},\ and\ \citenamefont {Tarucha}}]{Ito2018}%
  \BibitemOpen
  \bibfield  {author} {\bibinfo {author} {\bibfnamefont {T.}~\bibnamefont
  {Ito}}, \bibinfo {author} {\bibfnamefont {T.}~\bibnamefont {Otsuka}},
  \bibinfo {author} {\bibfnamefont {T.}~\bibnamefont {Nakajima}}, \bibinfo
  {author} {\bibfnamefont {M.~R.}\ \bibnamefont {Delbecq}}, \bibinfo {author}
  {\bibfnamefont {S.}~\bibnamefont {Amaha}}, \bibinfo {author} {\bibfnamefont
  {J.}~\bibnamefont {Yoneda}}, \bibinfo {author} {\bibfnamefont
  {K.}~\bibnamefont {Takeda}}, \bibinfo {author} {\bibfnamefont
  {A.}~\bibnamefont {Noiri}}, \bibinfo {author} {\bibfnamefont
  {G.}~\bibnamefont {Allison}}, \bibinfo {author} {\bibfnamefont
  {A.}~\bibnamefont {Ludwig}}, \bibinfo {author} {\bibfnamefont {A.~D.}\
  \bibnamefont {Wieck}}, \ and\ \bibinfo {author} {\bibfnamefont
  {S.}~\bibnamefont {Tarucha}},\ }\bibfield  {title} {\enquote {\bibinfo
  {title} {Four single-spin {R}abi oscillations in a quadruple quantum dot},}\
  }\href {\doibase 10.1063/1.5040280} {\bibfield  {journal} {\bibinfo
  {journal} {Appl. Phys. Lett.}\ }\textbf {\bibinfo {volume} {113}},\ \bibinfo
  {pages} {093102} (\bibinfo {year} {2018})}\BibitemShut {NoStop}%
\bibitem [{\citenamefont {{Mortemousque}}\ \emph {et~al.}(2018)\citenamefont
  {{Mortemousque}}, \citenamefont {{Chanrion}}, \citenamefont {{Jadot}},
  \citenamefont {{Flentje}}, \citenamefont {{Ludwig}}, \citenamefont {{Wieck}},
  \citenamefont {{Urdampilleta}}, \citenamefont {{Bauerle}},\ and\
  \citenamefont {{Meunier}}}]{Mortemousque2018}%
  \BibitemOpen
  \bibfield  {author} {\bibinfo {author} {\bibfnamefont {P.-A.}\ \bibnamefont
  {{Mortemousque}}}, \bibinfo {author} {\bibfnamefont {E.}~\bibnamefont
  {{Chanrion}}}, \bibinfo {author} {\bibfnamefont {B.}~\bibnamefont {{Jadot}}},
  \bibinfo {author} {\bibfnamefont {H.}~\bibnamefont {{Flentje}}}, \bibinfo
  {author} {\bibfnamefont {A.}~\bibnamefont {{Ludwig}}}, \bibinfo {author}
  {\bibfnamefont {A.~D.}\ \bibnamefont {{Wieck}}}, \bibinfo {author}
  {\bibfnamefont {M.}~\bibnamefont {{Urdampilleta}}}, \bibinfo {author}
  {\bibfnamefont {C.}~\bibnamefont {{Bauerle}}}, \ and\ \bibinfo {author}
  {\bibfnamefont {T.}~\bibnamefont {{Meunier}}},\ }\bibfield  {title} {\enquote
  {\bibinfo {title} {{Coherent control of individual electron spins in a two
  dimensional array of quantum dots}},}\ }\href@noop {} {\bibfield  {journal}
  {\bibinfo  {journal} {arXiv:1808.06180}\ } (\bibinfo {year}
  {2018})}\BibitemShut {NoStop}%
\bibitem [{\citenamefont {Mukhopadhyay}\ \emph {et~al.}(2018)\citenamefont
  {Mukhopadhyay}, \citenamefont {Dehollain}, \citenamefont {Reichl},
  \citenamefont {Wegscheider},\ and\ \citenamefont {Vandersypen}}]{Udi2018}%
  \BibitemOpen
  \bibfield  {author} {\bibinfo {author} {\bibfnamefont {U.}~\bibnamefont
  {Mukhopadhyay}}, \bibinfo {author} {\bibfnamefont {J.~P.}\ \bibnamefont
  {Dehollain}}, \bibinfo {author} {\bibfnamefont {C.}~\bibnamefont {Reichl}},
  \bibinfo {author} {\bibfnamefont {W.}~\bibnamefont {Wegscheider}}, \ and\
  \bibinfo {author} {\bibfnamefont {L.~M.~K.}\ \bibnamefont {Vandersypen}},\
  }\bibfield  {title} {\enquote {\bibinfo {title} {A 2x�2 quantum dot array
  with controllable inter-dot tunnel couplings},}\ }\href {\doibase
  10.1063/1.5025928} {\bibfield  {journal} {\bibinfo  {journal} {Appl. Phys.
  Lett.}\ }\textbf {\bibinfo {volume} {112}},\ \bibinfo {pages} {183505}
  (\bibinfo {year} {2018})}\BibitemShut {NoStop}%
\bibitem [{\citenamefont {Reed}\ \emph {et~al.}(2012)\citenamefont {Reed},
  \citenamefont {{DiCarlo}}, \citenamefont {Nigg}, \citenamefont {Sun},
  \citenamefont {Frunzio}, \citenamefont {Girvin},\ and\ \citenamefont
  {Schoelkopf}}]{Reedec}%
  \BibitemOpen
  \bibfield  {author} {\bibinfo {author} {\bibfnamefont {M.~D.}\ \bibnamefont
  {Reed}}, \bibinfo {author} {\bibfnamefont {L.}~\bibnamefont {{DiCarlo}}},
  \bibinfo {author} {\bibfnamefont {S.~E.}\ \bibnamefont {Nigg}}, \bibinfo
  {author} {\bibfnamefont {L.}~\bibnamefont {Sun}}, \bibinfo {author}
  {\bibfnamefont {L.}~\bibnamefont {Frunzio}}, \bibinfo {author} {\bibfnamefont
  {S.~M.}\ \bibnamefont {Girvin}}, \ and\ \bibinfo {author} {\bibfnamefont
  {R.~J.}\ \bibnamefont {Schoelkopf}},\ }\bibfield  {title} {\enquote {\bibinfo
  {title} {Realization of three-qubit quantum error correction with
  superconducting circuits},}\ }\href@noop {} {\bibfield  {journal} {\bibinfo
  {journal} {Nature}\ }\textbf {\bibinfo {volume} {482}},\ \bibinfo {pages}
  {382} (\bibinfo {year} {2012})}\BibitemShut {NoStop}%
\bibitem [{\citenamefont {Schindler}\ \emph {et~al.}(2011)\citenamefont
  {Schindler}, \citenamefont {Barreiro}, \citenamefont {Monz}, \citenamefont
  {Nebendahl}, \citenamefont {Nigg}, \citenamefont {Chwalla}, \citenamefont
  {Hennrich},\ and\ \citenamefont {Blatt}}]{Schindlerec}%
  \BibitemOpen
  \bibfield  {author} {\bibinfo {author} {\bibfnamefont {P.}~\bibnamefont
  {Schindler}}, \bibinfo {author} {\bibfnamefont {J.~T.}\ \bibnamefont
  {Barreiro}}, \bibinfo {author} {\bibfnamefont {T.}~\bibnamefont {Monz}},
  \bibinfo {author} {\bibfnamefont {V.}~\bibnamefont {Nebendahl}}, \bibinfo
  {author} {\bibfnamefont {D.}~\bibnamefont {Nigg}}, \bibinfo {author}
  {\bibfnamefont {M.}~\bibnamefont {Chwalla}}, \bibinfo {author} {\bibfnamefont
  {M.}~\bibnamefont {Hennrich}}, \ and\ \bibinfo {author} {\bibfnamefont
  {R.}~\bibnamefont {Blatt}},\ }\bibfield  {title} {\enquote {\bibinfo {title}
  {Experimental repetitive quantum error correction},}\ }\href {\doibase
  10.1126/science.1203329} {\bibfield  {journal} {\bibinfo  {journal}
  {Science}\ }\textbf {\bibinfo {volume} {332}},\ \bibinfo {pages} {1059}
  (\bibinfo {year} {2011})}\BibitemShut {NoStop}%
\bibitem [{\citenamefont {{Hensgens}}\ \emph {et~al.}(2017)\citenamefont
  {{Hensgens}}, \citenamefont {{Fujita}}, \citenamefont {{Janssen}},
  \citenamefont {{Li}}, \citenamefont {{van Diepen}}, \citenamefont {{Reichl}},
  \citenamefont {{Wegscheider}}, \citenamefont {{Das Sarma}},\ and\
  \citenamefont {{Vandersypen}}}]{Hensgens2017}%
  \BibitemOpen
  \bibfield  {author} {\bibinfo {author} {\bibfnamefont {T.}~\bibnamefont
  {{Hensgens}}}, \bibinfo {author} {\bibfnamefont {T.}~\bibnamefont
  {{Fujita}}}, \bibinfo {author} {\bibfnamefont {L.}~\bibnamefont {{Janssen}}},
  \bibinfo {author} {\bibfnamefont {X.}~\bibnamefont {{Li}}}, \bibinfo {author}
  {\bibfnamefont {C.~J.}\ \bibnamefont {{van Diepen}}}, \bibinfo {author}
  {\bibfnamefont {C.}~\bibnamefont {{Reichl}}}, \bibinfo {author}
  {\bibfnamefont {W.}~\bibnamefont {{Wegscheider}}}, \bibinfo {author}
  {\bibfnamefont {S.}~\bibnamefont {{Das Sarma}}}, \ and\ \bibinfo {author}
  {\bibfnamefont {L.~M.~K.}\ \bibnamefont {{Vandersypen}}},\ }\bibfield
  {title} {\enquote {\bibinfo {title} {{Quantum simulation of a Fermi-Hubbard
  model using a semiconductor quantum dot array}},}\ }\href {\doibase
  10.1038/nature23022} {\bibfield  {journal} {\bibinfo  {journal} {Nature}\
  }\textbf {\bibinfo {volume} {548}},\ \bibinfo {pages} {70} (\bibinfo {year}
  {2017})}\BibitemShut {NoStop}%
\bibitem [{\citenamefont {Georgescu}\ \emph {et~al.}(2014)\citenamefont
  {Georgescu}, \citenamefont {Ashhab},\ and\ \citenamefont
  {Nori}}]{GeorgescuQS}%
  \BibitemOpen
  \bibfield  {author} {\bibinfo {author} {\bibfnamefont {I.~M.}\ \bibnamefont
  {Georgescu}}, \bibinfo {author} {\bibfnamefont {S.}~\bibnamefont {Ashhab}}, \
  and\ \bibinfo {author} {\bibfnamefont {F.}~\bibnamefont {Nori}},\ }\bibfield
  {title} {\enquote {\bibinfo {title} {Quantum simulation},}\ }\href {\doibase
  10.1103/RevModPhys.86.153} {\bibfield  {journal} {\bibinfo  {journal} {Rev.
  Mod. Phys.}\ }\textbf {\bibinfo {volume} {86}},\ \bibinfo {pages} {153}
  (\bibinfo {year} {2014})}\BibitemShut {NoStop}%
\bibitem [{\citenamefont {Barthelemy}\ and\ \citenamefont
  {Vandersypen}(2013)}]{Barthelemyqs}%
  \BibitemOpen
  \bibfield  {author} {\bibinfo {author} {\bibfnamefont {P.}~\bibnamefont
  {Barthelemy}}\ and\ \bibinfo {author} {\bibfnamefont {L.~M.~K.}\ \bibnamefont
  {Vandersypen}},\ }\bibfield  {title} {\enquote {\bibinfo {title} {Quantum dot
  systems: a versatile platform for quantum simulations},}\ }\href {\doibase
  10.1002/andp.201300124} {\bibfield  {journal} {\bibinfo  {journal} {Ann.
  Phys.}\ }\textbf {\bibinfo {volume} {525}},\ \bibinfo {pages} {808} (\bibinfo
  {year} {2013})}\BibitemShut {NoStop}%
\bibitem [{\citenamefont {Byrnes}\ \emph {et~al.}(2008)\citenamefont {Byrnes},
  \citenamefont {Kim}, \citenamefont {Kusudo},\ and\ \citenamefont
  {Yamamoto}}]{Byrnesqs}%
  \BibitemOpen
  \bibfield  {author} {\bibinfo {author} {\bibfnamefont {T.}~\bibnamefont
  {Byrnes}}, \bibinfo {author} {\bibfnamefont {N.~Y.}\ \bibnamefont {Kim}},
  \bibinfo {author} {\bibfnamefont {K.}~\bibnamefont {Kusudo}}, \ and\ \bibinfo
  {author} {\bibfnamefont {Y.}~\bibnamefont {Yamamoto}},\ }\bibfield  {title}
  {\enquote {\bibinfo {title} {Quantum simulation of fermi-hubbard models in
  semiconductor quantum-dot arrays},}\ }\href {\doibase
  10.1103/PhysRevB.78.075320} {\bibfield  {journal} {\bibinfo  {journal} {Phys.
  Rev. B}\ }\textbf {\bibinfo {volume} {78}},\ \bibinfo {pages} {075320}
  (\bibinfo {year} {2008})}\BibitemShut {NoStop}%
\bibitem [{\citenamefont {Barnes}\ \emph {et~al.}(2019)\citenamefont {Barnes},
  \citenamefont {Nichol},\ and\ \citenamefont {Economou}}]{Barnestc}%
  \BibitemOpen
  \bibfield  {author} {\bibinfo {author} {\bibfnamefont {E.}~\bibnamefont
  {Barnes}}, \bibinfo {author} {\bibfnamefont {J.~M.}\ \bibnamefont {Nichol}},
  \ and\ \bibinfo {author} {\bibfnamefont {S.~E.}\ \bibnamefont {Economou}},\
  }\bibfield  {title} {\enquote {\bibinfo {title} {Stabilization and
  manipulation of multispin states in quantum-dot time crystals with heisenberg
  interactions},}\ }\href {\doibase 10.1103/PhysRevB.99.035311} {\bibfield
  {journal} {\bibinfo  {journal} {Phys. Rev. B}\ }\textbf {\bibinfo {volume}
  {99}},\ \bibinfo {pages} {035311} (\bibinfo {year} {2019})}\BibitemShut
  {NoStop}%
\bibitem [{\citenamefont {Petta}\ \emph {et~al.}(2005)\citenamefont {Petta},
  \citenamefont {Johnson}, \citenamefont {Taylor}, \citenamefont {Laird},
  \citenamefont {Yacoby}, \citenamefont {Lukin}, \citenamefont {Marcus},
  \citenamefont {Hanson},\ and\ \citenamefont {Gossard}}]{petta2005}%
  \BibitemOpen
  \bibfield  {author} {\bibinfo {author} {\bibfnamefont {J.~R.}\ \bibnamefont
  {Petta}}, \bibinfo {author} {\bibfnamefont {A.~C.}\ \bibnamefont {Johnson}},
  \bibinfo {author} {\bibfnamefont {J.~M.}\ \bibnamefont {Taylor}}, \bibinfo
  {author} {\bibfnamefont {E.~A.}\ \bibnamefont {Laird}}, \bibinfo {author}
  {\bibfnamefont {A.}~\bibnamefont {Yacoby}}, \bibinfo {author} {\bibfnamefont
  {M.~D.}\ \bibnamefont {Lukin}}, \bibinfo {author} {\bibfnamefont {C.~M.}\
  \bibnamefont {Marcus}}, \bibinfo {author} {\bibfnamefont {M.~P.}\
  \bibnamefont {Hanson}}, \ and\ \bibinfo {author} {\bibfnamefont {A.~C.}\
  \bibnamefont {Gossard}},\ }\bibfield  {title} {\enquote {\bibinfo {title}
  {Coherent manipulation of coupled electron spins in semiconductor quantum
  dots},}\ }\href {\doibase 10.1126/science.1116955} {\bibfield  {journal}
  {\bibinfo  {journal} {Science}\ }\textbf {\bibinfo {volume} {309}},\ \bibinfo
  {pages} {2180--2184} (\bibinfo {year} {2005})}\BibitemShut {NoStop}%
\bibitem [{\citenamefont {Russ}\ \emph {et~al.}(2018)\citenamefont {Russ},
  \citenamefont {Zajac}, \citenamefont {Sigillito}, \citenamefont {Borjans},
  \citenamefont {Taylor}, \citenamefont {Petta},\ and\ \citenamefont
  {Burkard}}]{Russ2018}%
  \BibitemOpen
  \bibfield  {author} {\bibinfo {author} {\bibfnamefont {M.}~\bibnamefont
  {Russ}}, \bibinfo {author} {\bibfnamefont {D.~M.}\ \bibnamefont {Zajac}},
  \bibinfo {author} {\bibfnamefont {A.~J.}\ \bibnamefont {Sigillito}}, \bibinfo
  {author} {\bibfnamefont {F.}~\bibnamefont {Borjans}}, \bibinfo {author}
  {\bibfnamefont {J.~M.}\ \bibnamefont {Taylor}}, \bibinfo {author}
  {\bibfnamefont {J.~R.}\ \bibnamefont {Petta}}, \ and\ \bibinfo {author}
  {\bibfnamefont {G.}~\bibnamefont {Burkard}},\ }\bibfield  {title} {\enquote
  {\bibinfo {title} {High-fidelity quantum gates in {Si/SiGe} double quantum
  dots},}\ }\href {\doibase 10.1103/PhysRevB.97.085421} {\bibfield  {journal}
  {\bibinfo  {journal} {Phys. Rev. B}\ }\textbf {\bibinfo {volume} {97}},\
  \bibinfo {pages} {085421} (\bibinfo {year} {2018})}\BibitemShut {NoStop}%
\bibitem [{\citenamefont {Zajac}\ \emph {et~al.}(2015)\citenamefont {Zajac},
  \citenamefont {Hazard}, \citenamefont {Mi}, \citenamefont {Wang},\ and\
  \citenamefont {Petta}}]{zajac2015reconfigurable}%
  \BibitemOpen
  \bibfield  {author} {\bibinfo {author} {\bibfnamefont {D.~M.}\ \bibnamefont
  {Zajac}}, \bibinfo {author} {\bibfnamefont {T.~M.}\ \bibnamefont {Hazard}},
  \bibinfo {author} {\bibfnamefont {X.}~\bibnamefont {Mi}}, \bibinfo {author}
  {\bibfnamefont {K.}~\bibnamefont {Wang}}, \ and\ \bibinfo {author}
  {\bibfnamefont {J.~R.}\ \bibnamefont {Petta}},\ }\bibfield  {title} {\enquote
  {\bibinfo {title} {A reconfigurable gate architecture for {Si/SiGe} quantum
  dots},}\ }\href@noop {} {\bibfield  {journal} {\bibinfo  {journal} {Appl.
  Phys. Lett.}\ }\textbf {\bibinfo {volume} {106}},\ \bibinfo {pages} {223507}
  (\bibinfo {year} {2015})}\BibitemShut {NoStop}%
\bibitem [{SOM()}]{SOM}%
  \BibitemOpen
  \href@noop {} {}\bibinfo {note} {See Supplemental Material at [URL will be
  inserted by publisher] for additional device characterization data including
  references
  \cite{Mills2018,Yoneda2015,ZajacScalable,Klauder1962,Witzel2010,Yoneda2018,Borjans2018}.}\BibitemShut
  {Stop}%
\bibitem [{\citenamefont {{Mills}}\ \emph {et~al.}(2019)\citenamefont
  {{Mills}}, \citenamefont {{Zajac}}, \citenamefont {{Gullans}}, \citenamefont
  {{Schupp}}, \citenamefont {{Hazard}},\ and\ \citenamefont
  {{Petta}}}]{Mills2018}%
  \BibitemOpen
  \bibfield  {author} {\bibinfo {author} {\bibfnamefont {A.~R.}\ \bibnamefont
  {{Mills}}}, \bibinfo {author} {\bibfnamefont {D.~M.}\ \bibnamefont
  {{Zajac}}}, \bibinfo {author} {\bibfnamefont {M.~J.}\ \bibnamefont
  {{Gullans}}}, \bibinfo {author} {\bibfnamefont {F.~J.}\ \bibnamefont
  {{Schupp}}}, \bibinfo {author} {\bibfnamefont {T.~M.}\ \bibnamefont
  {{Hazard}}}, \ and\ \bibinfo {author} {\bibfnamefont {J.~R.}\ \bibnamefont
  {{Petta}}},\ }\bibfield  {title} {\enquote {\bibinfo {title} {{Shuttling a
  single charge across a one-dimensional array of silicon quantum dots}},}\
  }\href@noop {} {\bibfield  {journal} {\bibinfo  {journal} {Nat. Commun.}\
  }\textbf {\bibinfo {volume} {10}},\ \bibinfo {pages} {1063} (\bibinfo {year}
  {2019})}\BibitemShut {NoStop}%
\bibitem [{\citenamefont {Baart}\ \emph {et~al.}(2016)\citenamefont {Baart},
  \citenamefont {Shafiei}, \citenamefont {Fujita}, \citenamefont {Reichl},
  \citenamefont {Wegscheider},\ and\ \citenamefont
  {Vandersypen}}]{baart2016single}%
  \BibitemOpen
  \bibfield  {author} {\bibinfo {author} {\bibfnamefont {T.~A.}\ \bibnamefont
  {Baart}}, \bibinfo {author} {\bibfnamefont {M.}~\bibnamefont {Shafiei}},
  \bibinfo {author} {\bibfnamefont {T.}~\bibnamefont {Fujita}}, \bibinfo
  {author} {\bibfnamefont {C.}~\bibnamefont {Reichl}}, \bibinfo {author}
  {\bibfnamefont {W.}~\bibnamefont {Wegscheider}}, \ and\ \bibinfo {author}
  {\bibfnamefont {L.~M.~K.}\ \bibnamefont {Vandersypen}},\ }\bibfield  {title}
  {\enquote {\bibinfo {title} {Single-spin {CCD}},}\ }\href@noop {} {\bibfield
  {journal} {\bibinfo  {journal} {Nat. Nanotechnol.}\ }\textbf {\bibinfo
  {volume} {11}},\ \bibinfo {pages} {330} (\bibinfo {year} {2016})}\BibitemShut
  {NoStop}%
\bibitem [{\citenamefont {van Diepen}\ \emph {et~al.}(2018)\citenamefont {van
  Diepen}, \citenamefont {Eendebak}, \citenamefont {Buijtendorp}, \citenamefont
  {Mukhopadhyay}, \citenamefont {Fujita}, \citenamefont {Reichl}, \citenamefont
  {Wegscheider},\ and\ \citenamefont {Vandersypen}}]{vanDiepen2018}%
  \BibitemOpen
  \bibfield  {author} {\bibinfo {author} {\bibfnamefont {C.~J.}\ \bibnamefont
  {van Diepen}}, \bibinfo {author} {\bibfnamefont {P.~T.}\ \bibnamefont
  {Eendebak}}, \bibinfo {author} {\bibfnamefont {B.~T.}\ \bibnamefont
  {Buijtendorp}}, \bibinfo {author} {\bibfnamefont {U.}~\bibnamefont
  {Mukhopadhyay}}, \bibinfo {author} {\bibfnamefont {T.}~\bibnamefont
  {Fujita}}, \bibinfo {author} {\bibfnamefont {C.}~\bibnamefont {Reichl}},
  \bibinfo {author} {\bibfnamefont {W.}~\bibnamefont {Wegscheider}}, \ and\
  \bibinfo {author} {\bibfnamefont {L.~M.~K.}\ \bibnamefont {Vandersypen}},\
  }\bibfield  {title} {\enquote {\bibinfo {title} {Automated tuning of
  inter-dot tunnel coupling in double quantum dots},}\ }\href {\doibase
  10.1063/1.5031034} {\bibfield  {journal} {\bibinfo  {journal} {Appl. Phys.
  Lett.}\ }\textbf {\bibinfo {volume} {113}},\ \bibinfo {pages} {033101}
  (\bibinfo {year} {2018})}\BibitemShut {NoStop}%
\bibitem [{\citenamefont {{Volk}}\ \emph {et~al.}(2019)\citenamefont {{Volk}},
  \citenamefont {{Zwerver}}, \citenamefont {{Mukhopadhyay}}, \citenamefont
  {{Eendebak}}, \citenamefont {{van Diepen}}, \citenamefont {{Dehollain}},
  \citenamefont {{Hensgens}}, \citenamefont {{Fujita}}, \citenamefont
  {{Reichl}}, \citenamefont {{Wegscheider}},\ and\ \citenamefont
  {{Vandersypen}}}]{Volk2019}%
  \BibitemOpen
  \bibfield  {author} {\bibinfo {author} {\bibfnamefont {C.}~\bibnamefont
  {{Volk}}}, \bibinfo {author} {\bibfnamefont {A.~M.~J.}\ \bibnamefont
  {{Zwerver}}}, \bibinfo {author} {\bibfnamefont {U.}~\bibnamefont
  {{Mukhopadhyay}}}, \bibinfo {author} {\bibfnamefont {P.~T.}\ \bibnamefont
  {{Eendebak}}}, \bibinfo {author} {\bibfnamefont {C.~J.}\ \bibnamefont {{van
  Diepen}}}, \bibinfo {author} {\bibfnamefont {J.~P.}\ \bibnamefont
  {{Dehollain}}}, \bibinfo {author} {\bibfnamefont {T.}~\bibnamefont
  {{Hensgens}}}, \bibinfo {author} {\bibfnamefont {T.}~\bibnamefont
  {{Fujita}}}, \bibinfo {author} {\bibfnamefont {C.}~\bibnamefont {{Reichl}}},
  \bibinfo {author} {\bibfnamefont {W.}~\bibnamefont {{Wegscheider}}}, \ and\
  \bibinfo {author} {\bibfnamefont {L.~M.~K.}\ \bibnamefont {{Vandersypen}}},\
  }\bibfield  {title} {\enquote {\bibinfo {title} {{Loading a quantum-dot based
  ``Qubyte'' register}},}\ }\href@noop {} {\bibfield  {journal} {\bibinfo
  {journal} {arXiv:1901.00426}\ } (\bibinfo {year} {2019})}\BibitemShut
  {NoStop}%
\bibitem [{\citenamefont {DiCarlo}\ \emph {et~al.}(2004)\citenamefont
  {DiCarlo}, \citenamefont {Lynch}, \citenamefont {Johnson}, \citenamefont
  {Childress}, \citenamefont {Crockett}, \citenamefont {Marcus}, \citenamefont
  {Hanson},\ and\ \citenamefont {Gossard}}]{dicarlo2004}%
  \BibitemOpen
  \bibfield  {author} {\bibinfo {author} {\bibfnamefont {L.}~\bibnamefont
  {DiCarlo}}, \bibinfo {author} {\bibfnamefont {H.~J.}\ \bibnamefont {Lynch}},
  \bibinfo {author} {\bibfnamefont {A.~C.}\ \bibnamefont {Johnson}}, \bibinfo
  {author} {\bibfnamefont {L.~I.}\ \bibnamefont {Childress}}, \bibinfo {author}
  {\bibfnamefont {K.}~\bibnamefont {Crockett}}, \bibinfo {author}
  {\bibfnamefont {C.~M.}\ \bibnamefont {Marcus}}, \bibinfo {author}
  {\bibfnamefont {M.~P.}\ \bibnamefont {Hanson}}, \ and\ \bibinfo {author}
  {\bibfnamefont {A.~C.}\ \bibnamefont {Gossard}},\ }\bibfield  {title}
  {\enquote {\bibinfo {title} {Differential charge sensing and charge
  delocalization in a tunable double quantum dot},}\ }\href {\doibase
  10.1103/PhysRevLett.92.226801} {\bibfield  {journal} {\bibinfo  {journal}
  {Phys. Rev. Lett.}\ }\textbf {\bibinfo {volume} {92}},\ \bibinfo {pages}
  {226801} (\bibinfo {year} {2004})}\BibitemShut {NoStop}%
\bibitem [{\citenamefont {Petta}\ \emph {et~al.}(2004)\citenamefont {Petta},
  \citenamefont {Johnson}, \citenamefont {Marcus}, \citenamefont {Hanson},\
  and\ \citenamefont {Gossard}}]{Petta2003}%
  \BibitemOpen
  \bibfield  {author} {\bibinfo {author} {\bibfnamefont {J.~R.}\ \bibnamefont
  {Petta}}, \bibinfo {author} {\bibfnamefont {A.~C.}\ \bibnamefont {Johnson}},
  \bibinfo {author} {\bibfnamefont {C.~M.}\ \bibnamefont {Marcus}}, \bibinfo
  {author} {\bibfnamefont {M.~P.}\ \bibnamefont {Hanson}}, \ and\ \bibinfo
  {author} {\bibfnamefont {A.~C.}\ \bibnamefont {Gossard}},\ }\bibfield
  {title} {\enquote {\bibinfo {title} {Manipulation of a single charge in a
  double quantum dot},}\ }\href {\doibase 10.1103/PhysRevLett.93.186802}
  {\bibfield  {journal} {\bibinfo  {journal} {Phys. Rev. Lett.}\ }\textbf
  {\bibinfo {volume} {93}},\ \bibinfo {pages} {186802} (\bibinfo {year}
  {2004})}\BibitemShut {NoStop}%
\bibitem [{\citenamefont {Pioro-Ladriere}\ \emph {et~al.}(2008)\citenamefont
  {Pioro-Ladriere}, \citenamefont {Obata}, \citenamefont {Tokura},
  \citenamefont {Shin}, \citenamefont {Kubo}, \citenamefont {Yoshida},
  \citenamefont {Taniyama},\ and\ \citenamefont
  {Tarucha}}]{pioro2008electrically}%
  \BibitemOpen
  \bibfield  {author} {\bibinfo {author} {\bibfnamefont {M.}~\bibnamefont
  {Pioro-Ladriere}}, \bibinfo {author} {\bibfnamefont {T.}~\bibnamefont
  {Obata}}, \bibinfo {author} {\bibfnamefont {Y.}~\bibnamefont {Tokura}},
  \bibinfo {author} {\bibfnamefont {Y.-S.}\ \bibnamefont {Shin}}, \bibinfo
  {author} {\bibfnamefont {T.}~\bibnamefont {Kubo}}, \bibinfo {author}
  {\bibfnamefont {K.}~\bibnamefont {Yoshida}}, \bibinfo {author} {\bibfnamefont
  {T.}~\bibnamefont {Taniyama}}, \ and\ \bibinfo {author} {\bibfnamefont
  {S.}~\bibnamefont {Tarucha}},\ }\bibfield  {title} {\enquote {\bibinfo
  {title} {Electrically driven single-electron spin resonance in a slanting
  {Z}eeman field},}\ }\href@noop {} {\bibfield  {journal} {\bibinfo  {journal}
  {Nat. Phys.}\ }\textbf {\bibinfo {volume} {4}},\ \bibinfo {pages} {776}
  (\bibinfo {year} {2008})}\BibitemShut {NoStop}%
\bibitem [{\citenamefont {Tokura}\ \emph {et~al.}(2006)\citenamefont {Tokura},
  \citenamefont {van~der Wiel}, \citenamefont {Obata},\ and\ \citenamefont
  {Tarucha}}]{Tokura2006}%
  \BibitemOpen
  \bibfield  {author} {\bibinfo {author} {\bibfnamefont {Y.}~\bibnamefont
  {Tokura}}, \bibinfo {author} {\bibfnamefont {W.~G.}\ \bibnamefont {van~der
  Wiel}}, \bibinfo {author} {\bibfnamefont {T.}~\bibnamefont {Obata}}, \ and\
  \bibinfo {author} {\bibfnamefont {S.}~\bibnamefont {Tarucha}},\ }\bibfield
  {title} {\enquote {\bibinfo {title} {Coherent single electron spin control in
  a slanting {Zeeman} field},}\ }\href {\doibase 10.1103/PhysRevLett.96.047202}
  {\bibfield  {journal} {\bibinfo  {journal} {Phys. Rev. Lett.}\ }\textbf
  {\bibinfo {volume} {96}},\ \bibinfo {pages} {047202} (\bibinfo {year}
  {2006})}\BibitemShut {NoStop}%
\bibitem [{\citenamefont {{Yoneda}}\ \emph {et~al.}(2015)\citenamefont
  {{Yoneda}}, \citenamefont {{Otsuka}}, \citenamefont {{Takakura}},
  \citenamefont {{Pioro- Ladri{\`e}re}}, \citenamefont {{Brunner}},
  \citenamefont {{Lu}}, \citenamefont {{Nakajima}}, \citenamefont {{Obata}},
  \citenamefont {{Noiri}}, \citenamefont {{Palmstr{\o}m}}, \citenamefont
  {{Gossard}},\ and\ \citenamefont {{Tarucha}}}]{Yoneda2015}%
  \BibitemOpen
  \bibfield  {author} {\bibinfo {author} {\bibfnamefont {J.}~\bibnamefont
  {{Yoneda}}}, \bibinfo {author} {\bibfnamefont {T.}~\bibnamefont {{Otsuka}}},
  \bibinfo {author} {\bibfnamefont {T.}~\bibnamefont {{Takakura}}}, \bibinfo
  {author} {\bibfnamefont {M.}~\bibnamefont {{Pioro- Ladri{\`e}re}}}, \bibinfo
  {author} {\bibfnamefont {R.}~\bibnamefont {{Brunner}}}, \bibinfo {author}
  {\bibfnamefont {H.}~\bibnamefont {{Lu}}}, \bibinfo {author} {\bibfnamefont
  {T.}~\bibnamefont {{Nakajima}}}, \bibinfo {author} {\bibfnamefont
  {T.}~\bibnamefont {{Obata}}}, \bibinfo {author} {\bibfnamefont
  {A.}~\bibnamefont {{Noiri}}}, \bibinfo {author} {\bibfnamefont {C.~J.}\
  \bibnamefont {{Palmstr{\o}m}}}, \bibinfo {author} {\bibfnamefont {A.~C.}\
  \bibnamefont {{Gossard}}}, \ and\ \bibinfo {author} {\bibfnamefont
  {S.}~\bibnamefont {{Tarucha}}},\ }\bibfield  {title} {\enquote {\bibinfo
  {title} {{Robust micromagnet design for fast electrical manipulations of
  single spins in quantum dots}},}\ }\href {\doibase 10.7567/APEX.8.084401}
  {\bibfield  {journal} {\bibinfo  {journal} {APEX}\ }\textbf {\bibinfo
  {volume} {8}},\ \bibinfo {eid} {084401} (\bibinfo {year} {2015})}\BibitemShut
  {NoStop}%
\bibitem [{\citenamefont {Elzerman}\ \emph {et~al.}(2004)\citenamefont
  {Elzerman}, \citenamefont {Hanson}, \citenamefont {Van~Beveren},
  \citenamefont {Witkamp}, \citenamefont {Vandersypen},\ and\ \citenamefont
  {Kouwenhoven}}]{elzerman2004single}%
  \BibitemOpen
  \bibfield  {author} {\bibinfo {author} {\bibfnamefont {J.~M.}\ \bibnamefont
  {Elzerman}}, \bibinfo {author} {\bibfnamefont {R.}~\bibnamefont {Hanson}},
  \bibinfo {author} {\bibfnamefont {L.~H.~W.}\ \bibnamefont {Van~Beveren}},
  \bibinfo {author} {\bibfnamefont {B.}~\bibnamefont {Witkamp}}, \bibinfo
  {author} {\bibfnamefont {L.~M.~K.}\ \bibnamefont {Vandersypen}}, \ and\
  \bibinfo {author} {\bibfnamefont {L.~P.}\ \bibnamefont {Kouwenhoven}},\
  }\bibfield  {title} {\enquote {\bibinfo {title} {Single-shot read-out of an
  individual electron spin in a quantum dot},}\ }\href@noop {} {\bibfield
  {journal} {\bibinfo  {journal} {Nature}\ }\textbf {\bibinfo {volume} {430}},\
  \bibinfo {pages} {431} (\bibinfo {year} {2004})}\BibitemShut {NoStop}%
\bibitem [{\citenamefont {Deelman}\ \emph {et~al.}(2016)\citenamefont
  {Deelman}, \citenamefont {Edge},\ and\ \citenamefont
  {Jackson}}]{Deelman2016}%
  \BibitemOpen
  \bibfield  {author} {\bibinfo {author} {\bibfnamefont {P.~W.}\ \bibnamefont
  {Deelman}}, \bibinfo {author} {\bibfnamefont {L.~F.}\ \bibnamefont {Edge}}, \
  and\ \bibinfo {author} {\bibfnamefont {C.~A.}\ \bibnamefont {Jackson}},\
  }\bibfield  {title} {\enquote {\bibinfo {title} {Metamorphic materials for
  quantum computing},}\ }\href {\doibase 10.1557/mrs.2016.28} {\bibfield
  {journal} {\bibinfo  {journal} {Mater. Res. Bull.}\ }\textbf {\bibinfo
  {volume} {41}},\ \bibinfo {pages} {224} (\bibinfo {year} {2016})}\BibitemShut
  {NoStop}%
\bibitem [{\citenamefont {Richardson}\ \emph {et~al.}(2017)\citenamefont
  {Richardson}, \citenamefont {Jackson}, \citenamefont {Edge},\ and\
  \citenamefont {Deelman}}]{Richardson2017}%
  \BibitemOpen
  \bibfield  {author} {\bibinfo {author} {\bibfnamefont {C.~J.~K.}\
  \bibnamefont {Richardson}}, \bibinfo {author} {\bibfnamefont {C.~A.}\
  \bibnamefont {Jackson}}, \bibinfo {author} {\bibfnamefont {L.~F.}\
  \bibnamefont {Edge}}, \ and\ \bibinfo {author} {\bibfnamefont {P.~W.}\
  \bibnamefont {Deelman}},\ }\bibfield  {title} {\enquote {\bibinfo {title}
  {High-resolution x-ray reflection {F}ourier analysis of metamorphic {Si/SiGe}
  quantum wells},}\ }\href {\doibase 10.1116/1.4978595} {\bibfield  {journal}
  {\bibinfo  {journal} {J. Vac. Sci. Technol. B}\ }\textbf {\bibinfo {volume}
  {35}},\ \bibinfo {pages} {02B113} (\bibinfo {year} {2017})}\BibitemShut
  {NoStop}%
\bibitem [{\citenamefont {Witzel}\ \emph {et~al.}(2012)\citenamefont {Witzel},
  \citenamefont {Carroll}, \citenamefont {Cywi\ifmmode~\acute{n}\else
  \'{n}\fi{}ski},\ and\ \citenamefont {Das~Sarma}}]{Witzel2012}%
  \BibitemOpen
  \bibfield  {author} {\bibinfo {author} {\bibfnamefont {W.~M.}\ \bibnamefont
  {Witzel}}, \bibinfo {author} {\bibfnamefont {M.~S.}\ \bibnamefont {Carroll}},
  \bibinfo {author} {\bibfnamefont {L.}~\bibnamefont
  {Cywi\ifmmode~\acute{n}\else \'{n}\fi{}ski}}, \ and\ \bibinfo {author}
  {\bibfnamefont {S.}~\bibnamefont {Das~Sarma}},\ }\bibfield  {title} {\enquote
  {\bibinfo {title} {Quantum decoherence of the central spin in a sparse system
  of dipolar coupled spins},}\ }\href {\doibase 10.1103/PhysRevB.86.035452}
  {\bibfield  {journal} {\bibinfo  {journal} {Phys. Rev. B}\ }\textbf {\bibinfo
  {volume} {86}},\ \bibinfo {pages} {035452} (\bibinfo {year}
  {2012})}\BibitemShut {NoStop}%
\bibitem [{\citenamefont {{Borjans}}\ \emph {et~al.}(2018)\citenamefont
  {{Borjans}}, \citenamefont {{Zajac}}, \citenamefont {{Hazard}},\ and\
  \citenamefont {{Petta}}}]{Borjans2018}%
  \BibitemOpen
  \bibfield  {author} {\bibinfo {author} {\bibfnamefont {F.}~\bibnamefont
  {{Borjans}}}, \bibinfo {author} {\bibfnamefont {D.~M.}\ \bibnamefont
  {{Zajac}}}, \bibinfo {author} {\bibfnamefont {T.~M.}\ \bibnamefont
  {{Hazard}}}, \ and\ \bibinfo {author} {\bibfnamefont {J.~R.}\ \bibnamefont
  {{Petta}}},\ }\bibfield  {title} {\enquote {\bibinfo {title} {Single-spin
  relaxation in a synthetic spin-orbit field},}\ }\href@noop {} {\bibfield
  {journal} {\bibinfo  {journal} {arXiv:1811.00848}\ } (\bibinfo {year}
  {2018})}\BibitemShut {NoStop}%
\bibitem [{\citenamefont {Klauder}\ and\ \citenamefont
  {Anderson}(1962)}]{Klauder1962}%
  \BibitemOpen
  \bibfield  {author} {\bibinfo {author} {\bibfnamefont {J.~R.}\ \bibnamefont
  {Klauder}}\ and\ \bibinfo {author} {\bibfnamefont {P.~W.}\ \bibnamefont
  {Anderson}},\ }\bibfield  {title} {\enquote {\bibinfo {title} {Spectral
  diffusion decay in spin resonance experiments},}\ }\href {\doibase
  10.1103/PhysRev.125.912} {\bibfield  {journal} {\bibinfo  {journal} {Phys.
  Rev.}\ }\textbf {\bibinfo {volume} {125}},\ \bibinfo {pages} {912--932}
  (\bibinfo {year} {1962})}\BibitemShut {NoStop}%
\bibitem [{\citenamefont {Witzel}\ \emph {et~al.}(2010)\citenamefont {Witzel},
  \citenamefont {Carroll}, \citenamefont {Morello}, \citenamefont
  {Cywi\ifmmode~\acute{n}\else \'{n}\fi{}ski},\ and\ \citenamefont
  {Das~Sarma}}]{Witzel2010}%
  \BibitemOpen
  \bibfield  {author} {\bibinfo {author} {\bibfnamefont {W.~M.}\ \bibnamefont
  {Witzel}}, \bibinfo {author} {\bibfnamefont {M.~S.}\ \bibnamefont {Carroll}},
  \bibinfo {author} {\bibfnamefont {A.}~\bibnamefont {Morello}}, \bibinfo
  {author} {\bibfnamefont {L.}~\bibnamefont {Cywi\ifmmode~\acute{n}\else
  \'{n}\fi{}ski}}, \ and\ \bibinfo {author} {\bibfnamefont {S.}~\bibnamefont
  {Das~Sarma}},\ }\bibfield  {title} {\enquote {\bibinfo {title} {Electron spin
  decoherence in isotope-enriched silicon},}\ }\href {\doibase
  10.1103/PhysRevLett.105.187602} {\bibfield  {journal} {\bibinfo  {journal}
  {Phys. Rev. Lett.}\ }\textbf {\bibinfo {volume} {105}},\ \bibinfo {pages}
  {187602} (\bibinfo {year} {2010})}\BibitemShut {NoStop}%
\end{thebibliography}

%

\end{document}